\documentclass{article} % For LaTeX2e
\usepackage{iclr2026_conference,times}
\iclrfinalcopy

\usepackage{amsmath,amsfonts,bm}

\def\eqref#1{equation~\ref{#1}}
\def\1{\bm{1}}

\DeclareMathAlphabet{\mathsfit}{\encodingdefault}{\sfdefault}{m}{sl}
\SetMathAlphabet{\mathsfit}{bold}{\encodingdefault}{\sfdefault}{bx}{n}

\usepackage[T1]{fontenc}    % use 8-bit T1 fonts

\usepackage{hyperref}       % hyperlinks
\usepackage{url}   
\usepackage{titlesec}
\usepackage{booktabs}       % professional-quality tables
\usepackage{amsfonts}       % blackboard math symbols
\usepackage{nicefrac}       % compact symbols for 1/2, etc.
\usepackage{microtype}      % microtypography
\usepackage{xcolor}         % colors
\usepackage{caption}
\usepackage{tabularray}
\usepackage[font=bf,labelfont=small]{caption}
\usepackage{wrapfig}

\usepackage{graphicx}
\usepackage{subcaption}

\usepackage{enumitem}
\usepackage{amsmath}
\usepackage{amssymb}
\usepackage{mathtools}
\usepackage{amsthm}
\usepackage{xspace}
\usepackage[most]{tcolorbox}
\usepackage{ulem}
\usepackage{listings}
\usepackage{hyperref}

\newcommand{\name}{{\textsc{NetArena}}\xspace}

\renewcommand{\cite}{\citep}

\title{\name:Dynamic Benchmarks for AI Agents in Network Automation}

\author{
  \textbf{Yajie Zhou}\textsuperscript{1} \quad
  \textbf{Jiajun Ruan}\textsuperscript{3} \quad
  \textbf{Eric S. Wang}\textsuperscript{1} \quad
  \textbf{Sadjad Fouladi}\textsuperscript{2}\\
  \textbf{\quad Francis Y. Yan}\textsuperscript{3} \quad
  \textbf{Kevin Hsieh}\textsuperscript{2} \quad
  \textbf{Zaoxing Liu}\textsuperscript{1} \\
  \textsuperscript{1}University of Maryland \quad
  \textsuperscript{2}Microsoft Research \quad
  \textsuperscript{3}University of Illinois Urbana-Champaign
}

\usepackage[most]{tcolorbox}
\usepackage{listings}
\usepackage[hashEnumerators,smartEllipses]{markdown}
\definecolor{promptaccent}{RGB}{63, 81, 181}       % Indigo accent
\definecolor{promptbg}{RGB}{250, 250, 252}         % Very light gray background
\definecolor{prompttitlebg}{RGB}{63, 81, 181}      % Indigo title background
\definecolor{keywordcolor}{RGB}{183, 28, 28}       % Deep red for emphasis
\definecolor{codeexcolor}{RGB}{27, 94, 32}         % Forest green for code
\definecolor{importantcolor}{RGB}{191, 54, 12}     % Burnt orange for important
\definecolor{labelcolor}{RGB}{69, 90, 100}         % Blue-gray for labels

\lstdefinestyle{outputstyle}{
    basicstyle=\small\ttfamily,
    breaklines=true,
    frame=none,
    backgroundcolor=\color{gray!5},
    xleftmargin=4pt,
    framexleftmargin=4pt,
}

\newcommand{\pkey}[1]{\textcolor{keywordcolor}{\textbf{#1}}}
\newcommand{\pcode}[1]{{\small\textcolor{codeexcolor}{\texttt{#1}}}}
\newcommand{\pimp}[1]{\textcolor{importantcolor}{\textbf{#1}}}
\newcommand{\plabel}[1]{\textcolor{labelcolor}{\textit{#1}}}

\newtcolorbox{PromptBlock}[1]{
    enhanced,
    breakable,
    colback=promptbg,
    colframe=promptaccent,
    coltitle=white,
    fonttitle=\bfseries\sffamily,
    title={#1},
    boxrule=0.5pt,
    arc=1pt,
    left=10pt,
    right=10pt,
    top=8pt,
    bottom=8pt,
    toptitle=4pt,
    bottomtitle=4pt,
    colbacktitle=prompttitlebg,
    sharp corners=downhill,
    borderline west={3pt}{0pt}{promptaccent}
    }

\titlespacing*{\section}
  {0pt}    % left indent
  {0.8ex plus 0.3ex minus 0.1ex}  % space before
  {0.6ex plus 0.2ex minus 0.1ex}  % space after

\titlespacing*{\subsection}
  {0pt}
  {0.7ex plus 0.2ex minus 0.1ex}
  {0.5ex plus 0.2ex minus 0.1ex}
\newcommand{\rev}[1]{{\color{black} #1}}

\begin{document}

\maketitle

\begin{abstract}
\label{sec:abstract}

As AI agents expand into high-stakes domains like network system operations, evaluating their real-world reliability becomes increasingly critical.
However, existing benchmarks risk contamination due to static design, show high statistical variance from limited dataset size, and fail to reflect the complexity of production environments.
We present \name, a dynamic benchmark generation framework for network applications. \name introduces a novel abstraction and unified interface that generalize across diverse tasks, enabling dynamic benchmarking despite the heterogeneity of network workloads. 
At runtime, users can generate unlimited queries on demand.
\name integrates with network emulators to measure correctness, safety, and latency during execution. 
We demonstrate \name on three representative applications and find that (1) \name significantly improves statistical reliability across AI agents, reducing confidence-interval overlap from 85\% to 0, 
(2) agents achieve only 13–38\% average performance (as low as 3\%) for large-scale, realistic queries, and (3) it exposes more fine-grained behaviors that static, correctness-only benchmarks miss.
% \name moves LLM evaluation toward realistic, scalable testing in real-world deployment. It enables use cases such as SFT and RL fine-tuning on network system tasks.
\name also enables use cases such as SFT and RL fine-tuning on network system tasks.
Code is available at~\url{https://github.com/Froot-NetSys/NetArena}.

\end{abstract}
\section{Introduction}
\label{sec:intro}

While LLMs~\cite{gpt-4o, gemini, llama3, yang2024qwen2} have rapidly advanced agent capabilities across general tasks~\cite{openai-cua,cua-survey}, many existing benchmarks focus on simplified settings that do not fully capture the demands of real-world deployments.
Network and system automation tasks offer a compelling alternative: they are high-stakes and require LLMs to reason under constraints like partial observability and operational risk.
From data center capacity planning~\cite{nemo_copilot} to root cause analysis~\cite{root-cause-1} and policy synthesis~\cite{wang2024netconfeval, sharma2023prosper}, these tasks require not just correctness, but robustness and efficiency, making them an ideal stress test for AI agents.

Despite the growing interest and advancements, rigorously evaluating AI agents in network and system operations remains an open challenge because of the \textbf{data scarcity problem}. These tasks often involve large-scale infrastructure with complex domain-specific logic, 
but current benchmarks rely on \textit{manually curated queries and ground truths} by domain experts.
This labor-intensive process has resulted in fewer than 300 queries in recent benchmarks~\cite{nemo_copilot, chen2025aiopslab} even after months of effort. 
Such small, static benchmarks introduce critical limitations: they are prone to statistical bias, vulnerable to data contamination~\cite{zhu2023dyval}, and raises concerns about generalizability. For instance, an agent that succeeds on one task may fail entirely when the topology, location, or workload shifts. Moreover, static datasets struggle to surface rare but important edge cases, which are crucial for robust evaluation yet infeasible to enumerate manually.

%in a computer network with thousands of hosts and tens of types of failure modes

%First, existing network benchmarks rely primarily on manual query construction, which is labor-intensive and difficult to scale. Recent work~\cite{nemo_copilot, chen2025aiopslab} includes fewer than 300 queries, handcrafted by network system domain experts over the course of more than one month. 
%Second, network applications operate over dynamic, high-dimensional state spaces. Statically enumerating all combinations of topologies, configurations, and failure modes is often cumbersome or even infeasible. For instance, in a routing network with over 1,000 hosts, each link may exhibit more than 20 types of error modes. As the network scales (e.g., to 2,000 hosts), the total number of possible error configurations grows exponentially, potentially reaching $20^{2000}$. 
%Static benchmarks in such cases are likely to miss critical edge cases. Moreover, static datasets are vulnerable to data contamination~\cite{zhu2023dyval}.

A natural way to address above challenges is to \textbf{dynamically generate queries} within the benchmark, as explored in recent work~\cite{zhu2023dyval, zhang2024darg, yu2024kieval}.
These methods often build symbolic graphs to synthesize diverse instances, mainly for arithmetic, logic, and program synthesis. However, they do not transfer well to network and system domains for two reasons.

First, unlike well-defined mathematical tasks, network and system problems rarely have a deterministic structure. This makes it hard to synthesize realistic queries and reliable ground truth~\cite{zhu2023dyval,zhang2024darg,yu2024kieval}. 
For example, troubleshooting a misconfiguration is not a one-step task. Agents must interact over multiple turns, collect diagnostics, infer root causes, and apply fixes in context. 
Second, success in operational tasks requires more than matching outputs to ground truth. Agents must avoid harmful side effects and respect system constraints such as safety and latency, which output-only evaluation often misses. Considering a network with thousands of hosts, one misconfigured command from AI agents can disable healthy paths and trigger cascading outages. 
Therefore, every agent action demands careful reasoning, and changes are only acceptable when both necessary and precisely scoped.

\begin{figure*}[t]
    \centering
    \includegraphics[width=0.99\linewidth]{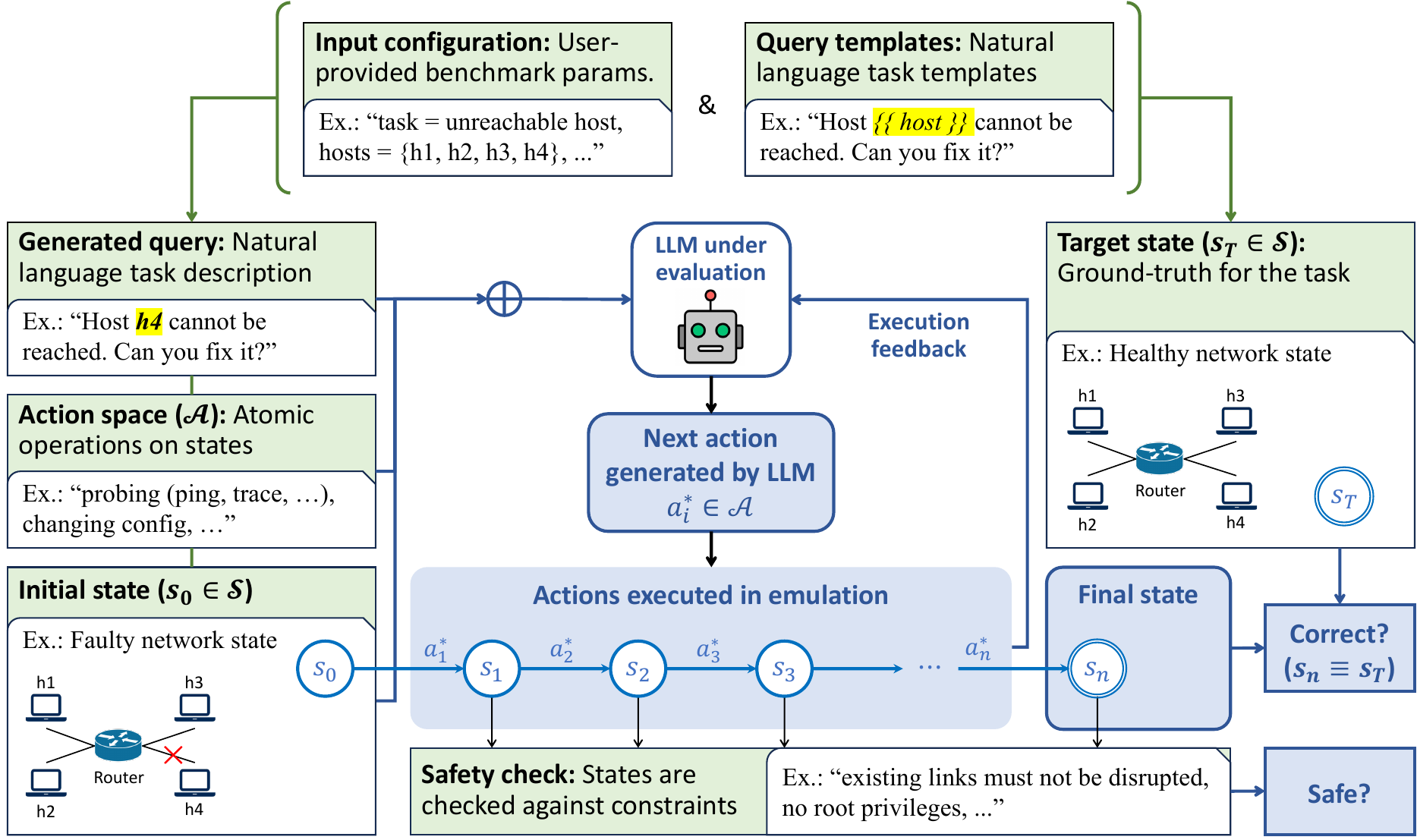}
    \caption{\name introduces a unified \textit{state-action} abstraction for a wide range of network and system applications, integrating with real-world emulators to generate dynamic queries and ground truths, and enabling automated correctness and safety evaluation.}
    \label{fig:design:e2e}
    % \vspace{-6mm}
\end{figure*}
%These characteristics highlight the need for a network-specific solution that supports dynamic query generation and realistic evaluation, incorporating multi-dimensional metrics. 

In this paper, we present \name, a novel framework for dynamic benchmark generation for real-world network automation tasks (Figure \ref{fig:design:e2e}). 
\name introduces a new evaluation paradigm, deploying agents in interactive, executable system environments to assess their capabilities through realistic, dynamically generated queries.
Our key technical contributions are as follows:
\begin{itemize}[topsep=1pt,itemsep=0ex,parsep=1ex, leftmargin=*]
    \item We define a \textbf{unified interface for abstracting network applications} based on explicit \textit{state} and \textit{action} spaces. This formalism supports dynamic query and ground truth generation (through executable state transitions), and enables controlled complexity scaling.
    
    \item By integrating with \rev{high-fidelity} network emulators (e.g., \citet{mininet}, Kubernetes \cite{google-k8s-demo}), \name enables \textbf{automatic, dynamic, and multi-turn verification} on LLM-generated actions, covering correctness, safety, and latency under \rev{deployment-like} conditions. 
    
    \item In \name, users only need to specify high-level configurations (e.g., query count, complexity, task type). \name \textbf{dynamically generates diverse evaluation sets} through stochastic sampling, which ensures broad coverage while reducing the risk of data contamination.
\end{itemize}

We instantiate \name in three representative network automation tasks: datacenter capacity planning, routing misconfiguration, and microservice policy troubleshooting. We evaluate five agents based on GPT-4o and QWen-72B models. Results on more LLMs and agents will appear on the project website~\footnote{Project website: \url{https://www.netarena.ai/}}.
% \todo{update link}.
Our key findings are:
\begin{itemize}[topsep=1pt,itemsep=0ex,parsep=1ex, leftmargin=*]
    \item \textbf{\rev{Agent performance is strikingly low.}} Average correctness across tasks is only 24\%, and even the best agent stays below 60\%. Small benchmarks ($<$200 queries) show high variance (average correctness rises to 38\%), making statistical comparisons unreliable. \name enables automated large-scale evaluation (e.g., $>$4000 queries), reducing confidence interval overlap between agents from as high as 85\% to 0\%, significantly improving evaluation reliability (\S\ref{eval:confidence_interval}).
    % \kevin{Give some high-level numbers on how the variance/overlap is reduced} \lesley{updated, plz check}

    \item \textbf{\rev{Correctness alone is insufficient.}} With metrics spanning correctness, safety, and latency, \name exposes tradeoffs in agents' behavior. Some models produce correct answers that violate system constraints and are unsafe, while others act conservatively, preserving safety but failing to resolve issues within acceptable latency (\S\ref{eval:spider_chart}).

    \item \textbf{\rev{Supervised fine-tuning (SFT) behaves inconsistently.}} For correctness, SFT models often overfit to the complexity of their training data, and only the model trained on data spanning all levels of task difficulty generalizes well. For safety, the simplest-level SFT model surprisingly generalizes best across all tasks, outperforming those trained on harder levels. With controlled variation of task difficulty and multi-dimensional metrics, \name enables such fine-grained analysis (\S\ref{eval:sft_bar}).
\end{itemize}
Beyond evaluation, we discuss future use cases for \name. First, we can add reward models in \name for on-policy reinforcement learning (RL)  fine-tuning. Second, \name can generate targeted adversarial queries to probe model weaknesses with rare but important corner cases (\S\ref{sec:use_case}).
% The detailed usage of \name is available at \url{https://github.com/Froot-NetSys/NetPress}.

% \begin{figure*}[t]
%     \centering
%      \begin{subfigure}[t]{0.9\textwidth}
%     {
%         \includegraphics[width=0.99\linewidth]{figs/benchmark_serving_stage.png}
%         \caption{Dynamic benchmark serving workflow (provide template to show dynamic?, multi-in parallel)}
%         \label{fig:serve}
%     }
%     \end{subfigure}
%     \begin{subfigure}[t]{0.9\textwidth}
%     {
%         \includegraphics[width=0.99\linewidth]{figs/benchmark_design_stage.png}
%         \caption{Breakdown of benchmark generation part (2) in the above workflow (zoom in, empha two examples. does not show how? Expert input for a and b, just one step. clear two users)}
%         \label{fig:design}

%     }
%     \end{subfigure}
%     \caption{Design of \name. }
%     \label{fig:design:e2e}
% \end{figure*}

\section{Related Work} 
% \kevin{For each category, we should discuss why they cannot address our need for systems/networking benchmarks} \lesley{added in the end of each category.}
% \lesley{do we need to cite much more paper here? I saw some NeurIPS paper have very long citation list}
\textbf{General Benchmarks for Evaluating LLMs.}
A growing body of work has focused on benchmarking the reasoning and autonomy of LLM agents~\cite{nathani2025mlgym, 
chang2024survey, gao2023adaptive, ribeiro2022adaptive, kiela2021dynabench, ma2021dynaboard, lambert2024rewardbench, li2024treeeval, li2023alpacaeval, lei2023s3eval, liang2022holistic, wang2024benchmark, yu2023skillmix, yu2024kieval, zhong2023agieval, huang2024ceval, hendrycks2020measuring, hendrycks2021math, starace2025paperbench, superbench}. 
  For example, CORE-Bench~\cite{corebench} aggregates reproducibility tasks from 90 papers to test agents’ ability to rerun experiments.
  % UPER-Bench~\cite{superbench} benchmarks agents on configuring and executing code from open-source ML/NLP projects of varying complexity. 
  RE-Bench~\cite{re-bench} compares agent solutions on open-ended ML research tasks to those from expert engineers. 
  MLE-Bench~\cite{mle-bench} converts 75 Kaggle competitions into agent benchmarks for leaderboard-based ML engineering. 
  SWE-Bench~\cite{swe-bench} requires agents to resolve GitHub issues. 
  % PaperBench~\cite{starace2025paperbench} tasks agents with reproducing 20 ICML 2024 papers. 
    Although these benchmarks are valuable in their own domains, they do not capture realistic network tasks that require deployment level reliability.
    
    \textbf{LLM Evaluation in Network and System Domains.}
In networking, LLMs have been used to generate graph-based network code~\cite{nemo_copilot,zhou2025meshagent}, synthesize configuration files~\cite{wang2024netconfeval}, support fault localization and remediation~\cite{root-cause-2}, and investigate internet incidents~\cite{zhou2023towards}. LLMs have also been applied to extract protocol specifications~\cite{sharma2023prosper}, reproduce networking experiments~\cite{xiang2023toward, kotaru2023adapting}, and evaluate system operations~\cite{chen2025aiopslab, jha2025itbench}. 
Beyond networking, AIOpsLab~\cite{chen2025aiopslab} introduces 48 tasks for assessing agent performance in DevOps scenarios. WebVoyager~\cite{he2024webvoyager} and WebArena~\cite{zhou2023webarena} benchmark LLM agents through real-world website interaction tasks, while OSWorld~\cite{xie2024osworld} evaluates 369 operating system tasks. The Berkeley Function-Calling Leaderboard (BFCL)~\cite{bfcl} measures agents' ability to correctly invoke APIs.
Existing benchmarks in these domains are static and rely on expert-driven manual curation, which constrains their scalability and raises significant concerns regarding data contamination (detailed comparisons in Table \ref{tab:design:compare_datset}).
\begin{table}
\centering
\resizebox{1\columnwidth}{!}{
\begin{tblr}{
  cells = {c},
  hline{1-2,6} = {-}{},
  hline{5} = {dashed}, % dashed line before last row
}
\textbf{Benchmark} & \textbf{Scale} & \textbf{Correctness(95\% CI)} & \textbf{Safety/Latency} & \textbf{Contamination Risk} & \textbf{Generalizability} \\

NeMoCopilot  & 33   & 94\%~[-]      & N/A & High (Static) & Low (Management) \\

AI4OpsLab    & 48   & 59\%~[-]      & N/A  & High (Static) & Low (DevOps) \\

NetConfEval  & 3200 & 100\%~[-]     & N/A  & High (Static) & Low (Configuration) \\

{\textbf{\name}\\\textbf{(Ours)}} & {9,250\\(unlimited)} & {44\%\\{[}0.01,~0.14]} & 35\% / 18s & Low (Dynamic) & {High (Management, \\K8s, Routing, etc)} \\
\end{tblr}

}
\caption{\name dynamically generates unlimited size of benchmarks for diverse network automation tasks, supporting more scalable and robust evaluation than existing benchmarks.}
\label{tab:design:compare_datset}
\vspace{-2mm}
\end{table}

\textbf{Dynamic Benchmark Generation.}
A line of work aims to reduce benchmark contamination risk~\cite{choi-etal-2024-unigen,balloccu2024leak, bender2021dangers, chen2021evaluating, deng2023investigating, dong2024generalization, golchin2023time, jacovi2023stop, jiang2024investigating, li2024latesteval, li2023estimating, li2024task, oren2023proving, roberts2023cutoff, sainz2023nlp, shi2023detecting, zhang2024careful, zhou2023dont, li2025treeeval, sun2024scieval, zhang2024darg}. DyVal~\cite{zhu2023dyval} introduces agents to generate and judge cognitively diverse variations of reasoning tasks. 
%DARG~\cite{zhang2024darg} perturbs symbolic reasoning graphs to control task complexity. 
KIEval~\cite{yu2024kieval} dynamically conducts multi-turn knowledge-based interactions. LatestEval~\cite{li2024latesteval} constructs test sets from new published material to avoid overlap with model pretraining. 
Dysca~\cite{zhang2024dysca} dynamically evaluates Vision LLMs by image synthesis.
%SciEval~\cite{sun2024scieval} targets scientific understanding through multi-skill question generation. 
%TreeEval~\cite{li2025treeeval} uses tree-structured plans to adaptively generate questions based on prior model responses. 
These dynamic benchmarks focus on general reasoning tasks, but are difficult to adapt to networking, where ground truth depends on system execution and cannot be reliably auto-generated.
%\lesley{While these dynamic benchmarks focus primarily on general reasoning tasks, they are difficult to extend to the networking domain, where ground truth often depends on real system execution and cannot be pre-generated reliably.}

% \kevin{Need to add another line of work on LLM evaluation for web or computer-use agents, such as WebArena, WebVoyage, OSWorld, etc. The function calling benchmark like Berkley Function Calling may be relevant here too} \lesley{Merged these into the second category.}

% Compared to prior work, \name targets a unique and underexplored challenge: dynamically generating benchmarks grounded in realistic network environments. 
% By incorporating feedback from network emulation, \name aim to bridge a critical gap in evaluating LLMs for infrastructure-scale applications.

% \input{sections/application_v2}
\section{\name}
In this section, we summarize two representative network tasks that can potentially be automated via LLM-based agents, present a unified pipeline for dynamic query and ground truth generation, and describe how \name automatically verifies each step of an agent’s output (Figure \ref{fig:design:e2e}).
% First fig missing input and output.
% need to talk about actual query (base prompt, golden answer).
% Bechmark application developer, we provide both prompt + system.
% Make it clear which part is ours and which is not.
% How the prompt and GT is generated

% Clarify what is done by user/what is done by system: user, agent, our system.
% (user provide a bunch of agents as input)

% (branch after actions, on how do we present the task in 2 ways: 1 is order, the delta. 2 is reverted, init state of generation become the targets. State, Action, Real Query to the model, goal are taken from different state. Expansion of (ii))

% Figure 3: remove redundancy.

% Start with How system works with 1234 (input and output)
% The fig part does not have too much info.

% Our goal: a set of app are promising for LLMs, but need a new way of evaluating it.

% Generating ground truth:
% 1. if we can break it down, we use DyVal-like approach, and can filter it with dependency graph.
% 2. if more complicated, we rely on the simulator to check the status.

% Discussion point:
% Mention that simulator could cause latency issue, we use all tricks to speed it up.
% Suggestion for new applications.
% Mention the difference (original VS now)
% How much throughput do we increase by doing so.

% Main:
% new application, how does each application work in details
% the component module.

% Why and how is it new?

% Capabilities.
% If the build up is not novel (engineering), then no focus, we.

% Claim the network topology (original state, the new state).

% Conceptually what's the difference between us and DyVal.
\begin{figure*}[t]
    \centering
    \begin{subfigure}[t]{0.49\textwidth}
    {
        \includegraphics[width=0.99\linewidth]{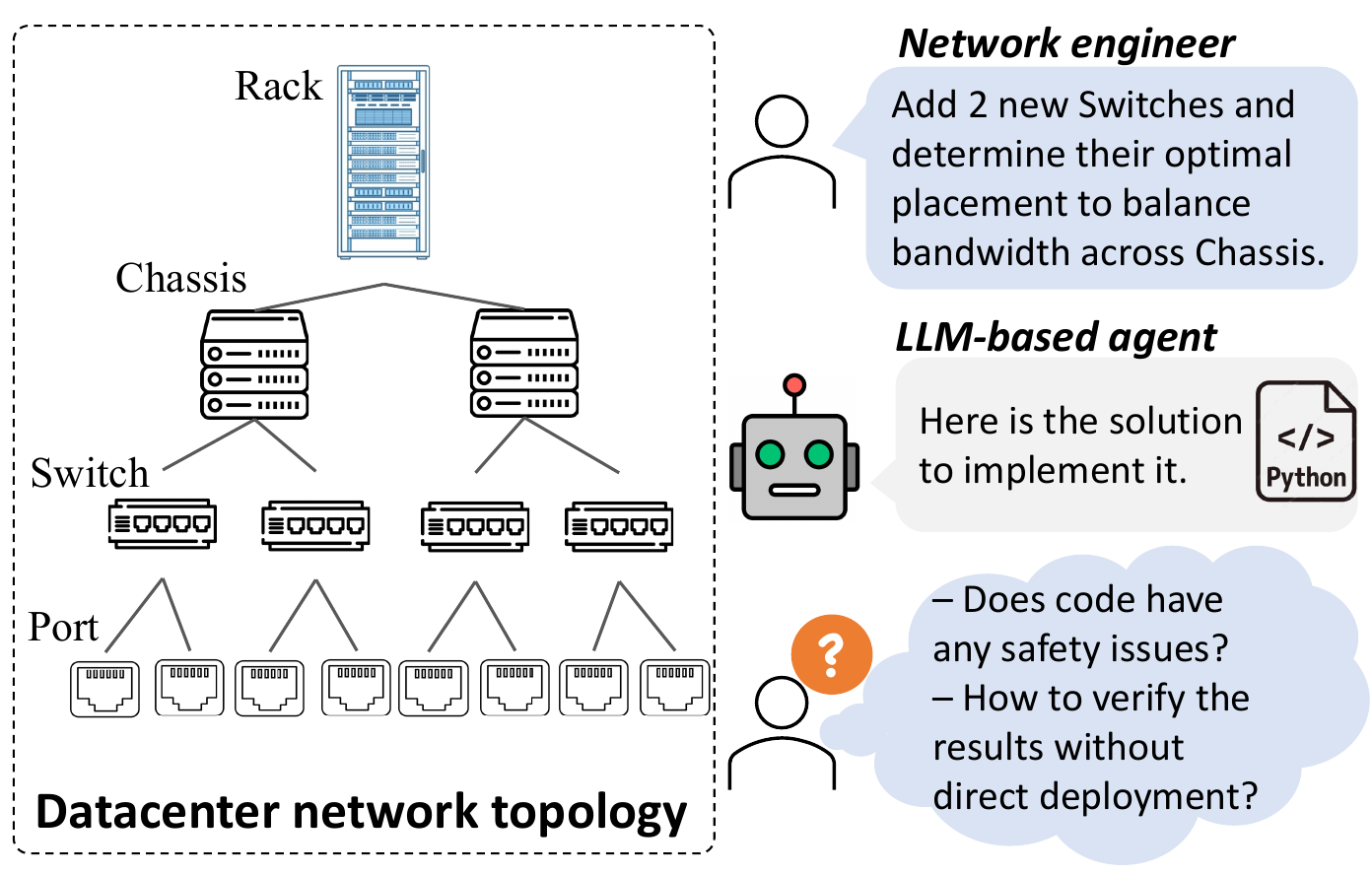}
        \caption{Constructive: datacenter capacity planning}
        \label{fig:malt}

    }
    \end{subfigure}
    \begin{subfigure}[t]{0.49\textwidth}
    {
        \includegraphics[width=0.99\linewidth]{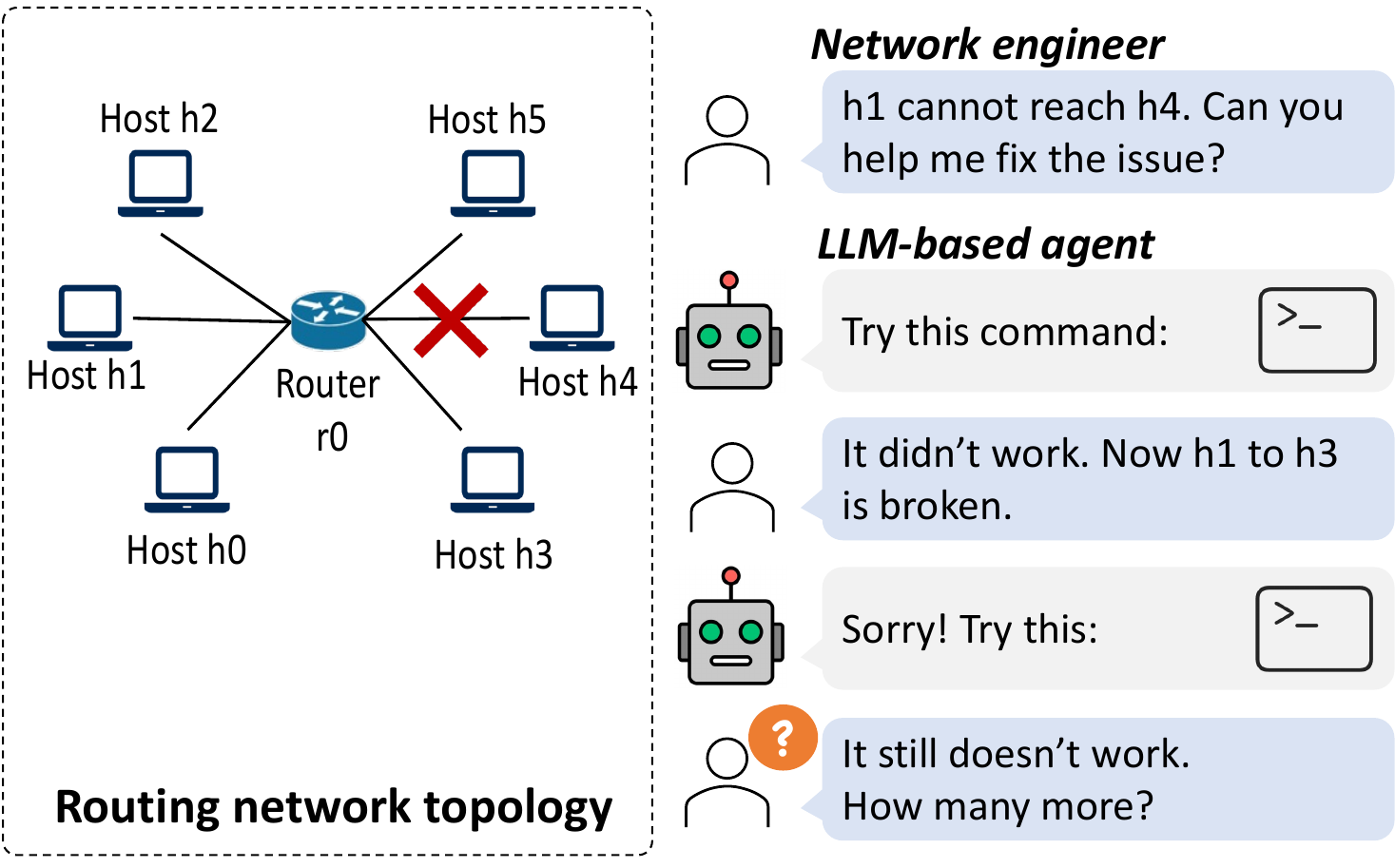}
        \caption{Reactive: routing misconfiguration}
        \label{fig:route}
    }
    \end{subfigure}
    \caption{Real-world network automation examples. }
    \label{fig:overview:example}
    \vspace{-2mm}
\end{figure*}

\subsection{LLM-based Network Automation Tasks} 
\label{sec:design:task_example}
% \kevin{The section title is about challenges, but this section is more about the types of applications. Consider re-framing this section by first introducing these applications and then say they can be grouped into two types.} \lesley{updated title, this is more like a background part}\kevin{Align the terminology across the paper (either "AIOps" or "systems and networking")}

Real-world network automation tasks require distinct forms of interaction between AI agents and the task environment. We highlight two representative classes of these tasks: 
% \kevin{Put a forward pointer to the appendix, and see if you can pull some of the appendix material here to clarify the tasks (prompt, goal, etc.)} \lesley{added pointer below, is it enough?}

\begin{itemize}[topsep=1pt,itemsep=0ex,parsep=1ex, leftmargin=*]

    \item \textbf{Constructive tasks} require structured solutions to well-specified queries. These resemble ``white-box'' settings where the query expresses a clear intent, and agents must generate a policy update that fulfills that intent. In such cases, LLM agents modify the current network state into a target state using interpretable operations. For example, in datacenter capacity planning (Figure~\ref{fig:malt}), users may request to find the optimal placement for new switches. The agent must synthesize a valid solution by composing various operations (e.g., \texttt{add}, \texttt{rank}, \texttt{update}). Although these tasks have deterministic outcomes, verifying correctness is challenging. It requires operational checks to ensure the solution adheres to policy constraints (e.g., the bandwidth must be over a minimum threshold). Appendix \ref{prompt:malt} describes a full example.

    \item \textbf{Reactive tasks} involve diagnosing faults and issuing repairs in under-specified, evolving environments. These resemble ``black-box'' settings, where the query specifies the problem but it is unclear how to fix it. In such cases, the LLM agent must iteratively observe, hypothesize, and act to identify the correct solution. For example, in a routing misconfiguration task (Figure~\ref{fig:route}), users may state, ``this link is down, help me fix it.'' The agent must perform multiple steps of information gathering and action, such as inspecting interface states, identifying missing routes, and applying targeted fixes. Since these tasks involve multi-turn interaction, evaluation cannot rely on predefined solutions but must assess whether the intended outcome (e.g., restored connectivity) is achieved without introducing new risks at each turn. Appendix \ref{prompt:route} provides a detailed example. 
\end{itemize}

% These real-world network application tasks go beyond static input-output matching. They require agents to reason over multiple steps, and anticipate side effects. 
% Scalable evaluation in such settings requires principles to model tasks and systematically generate benchmarks grounded in real execution environments.

\subsection{A Unified Abstraction for Generating Network Benchmarks}
\label{sec:design:generation}

The above real-world network automation tasks go beyond static input-output matching, making scalable evaluation far more complex. Effective benchmarking requires principled task modeling that enables systematic generation of queries and ground truths.
\rev{To address this, we define a unified pipeline with a general abstraction that consistently generates diverse queries and ground truths across applications under a single evaluation strategy.}
% Let:
% \begin{itemize}
%     \item $\mathcal{S}_\text{init}$: the initial network state defined by the application.
%     \item $\mathcal{A} = \{ a_1, a_2, \dots, a_k \}$: the set of \textit{basic actions} defined by application experts. These actions include valid state-changing operations (e.g., add, update, delete, inject error).
%     \item $\mathcal{Q} = \{ Q_1, Q_2, \dots, Q_n \}$: the set of dynamically generated queries, each composed from a combination of actions in $\mathcal{A}$.
%     \item $\mathcal{S}_{Q_i}$: the resulting target network state corresponding to query $Q_i$, computed by applying the basic actions of $Q_i$ to $\mathcal{S}_\text{init}$.
%     \item $\mathcal{P}_{\text{LLM}}(Q_i)$: the program or policy generated by the LLM for query $Q_i$, in the form of code or executable commands.
%     \item $\mathcal{S}_{\text{LLM}_i}$: the network state obtained after applying $\text{Ans}_{\text{LLM}}(Q_i)$ to $\mathcal{S}_\text{init}$.
%     \item $\mathcal{C}$: a set of system-level constraints or safety properties that must be preserved in all valid network states.
% \end{itemize}

\textbf{State Transition Process.}
While network automation tasks differ in detailed objectives, they often share a foundational structure: they operate over an underlying network/system topology (graph), and each step of the interaction involves analyzing or modifying the state of this topology. 
Each task's objective can be modeled as a finite state transition system $(\mathcal{S}, \mathcal{A}, \mathcal{E})$, where $\mathcal{S}$ is the set of system states, $\mathcal{A}$ the set of atomic action functions, and $\mathcal{E}$ the application-specific execution function. Each action $a_t \in \mathcal{A}$ is parameterized by task-specific operands $\theta_t$ and represented as $a_t(\theta_t)$.

A benchmark query defines an execution episode that begins from an initial state $s_0 \in \mathcal{S}$ and applies a sequence of $T$ such parameterized actions $\{a_0(\theta_0), a_1(\theta_1), \ldots, a_{T-1}(\theta_{T-1})\}$:
\begin{equation}
    s_{t+1} = \mathcal{E}(s_t, a_t(\theta_t)), \quad \text{for } t = 0, \ldots, T-1
\end{equation}
To instantiate a new task in \name, developers only need to define the state space $\mathcal{S}$ (e.g., a routing topology with connectivity status) and the action space $\mathcal{A}$ (e.g., \texttt{IP(u)} indicating a link-level IP error, where \texttt{u} is the operand link). 

\textbf{Query and Ground Truth Generation.} We distinguish between the two types of tasks based on how queries and their corresponding ground truths are generated:

\textit{(1) Constructive tasks.} These tasks start from a known initial state $s_{\text{init}}$, and the goal is to generate a sequence of actions that deterministically transition the system to a specified target state $s_T$. The ground truth is given by a predefined action sequence $\mathcal{A}^* = \{a_0^*, a_1^*, \ldots, a_{T-1}^*\}$, which yields $s_T$ under the execution function:
\begin{equation}
    \mathcal{E}(s_0, \mathcal{A}^*) \triangleq \left(\mathcal{E}(\cdot, a_{T-1}^*(\theta_{T-1}^*)) \circ \cdots \circ \mathcal{E}(\cdot, a_1^*(\theta_1^*)) \circ \mathcal{E}(\cdot, a_0^*(\theta_0^*)) \right)(s_0) = s_T
\end{equation}
Here, $\mathcal{E}_{a(\theta)}$ denotes applying a parameterized action to a state, and $\circ$ represents functional composition. This captures the cumulative transformation of the system through sequential actions.

When generating a new query, \name samples an initial state $s_0$ and a set of action $\mathcal{A}^*$, with each action's operand dynamically drawn from a large space. By executing these actions on $s_0$, \name produces the intended goal state $s_T$. A natural language template then converts $(s_0, \mathcal{A}^*, s_T)$ into a prompt for the LLM agent. The agent’s correctness is evaluated by comparing its resulting state to $s_T$, and, when available, its predicted action sequence to $\mathcal{A}^*$ to assess the agent's reasoning process.

% When generating a new query, \name samples an initial state $s_0$ and a ground-truth action sequence $\mathcal{A}^$, with each action’s operand dynamically drawn from a large space. Executing $\mathcal{A}^$ on $s_0$ yields the goal state $s_T$. A natural language template then converts $(s_0, \mathcal{A}^*, s_T)$ into a prompt for the LLM agent. The agent’s correctness is evaluated by comparing its resulting state to $s_T$, and, when available, its predicted action sequence to $\mathcal{A}^*$.

% \lesley{expand this paragraph to explain how the queries, states, and action sequences are generated} LLM agents receive queries as natural language descriptions of the initial state $s_0$, along with high-level instructions describing the desired transformation to the intended goal state $s_T$. 
% (e.g., a capacity planning query may state: ``Given the topology, add four new Packet Switches and balance the capacity load across Chassis.'')
% The correctness of the LLM's output is evaluated by comparing its resulting state to the target state $s_T$ (i.e., ground truth state). When available, the intermediate action sequence can also be checked against the ground truth $\mathcal{A}^*$ to assess its reasoning steps.

\textit{(2) Reactive tasks.} These tasks begin from a faulty network state $s_{\text{faulty}}$, generated by applying a hidden fault injection sequence $\mathcal{A}_{\text{inj}} = \{a_0^{\text{inj}}(\theta_0^{\text{inj}}), \ldots, a_K^{\text{inj}}(\theta_K^{\text{inj}})\}$ to an originally healthy state $s_0$:
\begin{equation}
s_{\text{faulty}} = \mathcal{E}(s_0, \mathcal{A}_{\text{inj}})
\end{equation}
Here, the fault injection sequence $\mathcal{A}_{\text{inj}}$ is hidden from the LLM agent. 

When generating a new query, \name injects errors into the original healthy state $s_0$, producing a faulty state $s_T$ that serves as the query input. A natural language template then describes $s_T$, and the LLM agent is tasked with recovering the system to $s_0$. Unlike constructive tasks, multiple valid recovery paths may exist for a single faulty state. Correctness is therefore judged by whether the agent restores the system to $s_0$, rather than by matching the specific injected action sequence $\mathcal{A}_{\text{inj}}$.

% The agent must generate a recovery sequence $\hat{\mathcal{A}} = \{\hat{a}_0(\hat{\theta}_0), \ldots, \hat{a}_{T-1}(\hat{\theta}_{T-1})\}$ that successfully restores the system.

% \lesley{expand this paragraph to explain how the queries, states, and fault injection sequences are generated} Unlike constructive tasks, multiple valid recovery paths may exist for one faulty state. Therefore, correctness is evaluated based on the equivalence of the final recovered state to the original healthy state $s_0$ (i.e., ground truth state), rather than alignment with a specific action trace. For example, a routing troubleshooting query may state: ``Host-A cannot reach Router-0, diagnose and fix the issue.'' The agent must issue diagnostic and corrective commands that ultimately re-establish connectivity. 

%Note that the recovery sequence $\hat{\mathcal{A}}$ often involves probing and hypothesizing, and there are many sequences that can achieve the correct healthy state.
%Since recovery strategies may differ, no single ground truth sequence is enforced.

\subsection{Realistic Agent Evaluation via Emulator Integration}
\label{sec:design:metric}

A key challenge in deploying AI agents for network automation tasks is the uncertainty of their real-world behavior, including side effects, security risks, and inefficiencies. Testing these behaviors in real production environments is infeasible due to risk and cost \cite{llm_industry_risk}. To address this, \name integrates directly with high-fidelity network emulators, providing controlled and reproducible evaluation under realistic conditions with diverse system metrics.

% \textbf{Integration with Emulators.}
% We embed agent evaluation within realistic execution environments to closely mimic real deployment scenarios. Agent actions are executed end-to-end, and their effects are validated using feedback from the environment. For example, in a routing application, the Mininet \cite{mininet} simulator can verify whether connectivity is successfully restored and whether the agent’s commands introduce new risks—such as inadvertently disabling previously functional links.

\textbf{Comprehensive Performance Metrics from Emulator Integration.}
We embed agent evaluation in high-fidelity network emulators, \rev{which provide the closest deployment-like environments without the risks of production. These emulators are widely adopted in both academia and industry (e.g., Google, Alibaba, etc) for testing large-scale systems.} Agent actions are executed end-to-end, and their effects are validated through emulator feedback. For instance, in a routing task, ~\citet{mininet} can verify whether connectivity is restored and whether new risks emerge, such as inadvertently disabling functional links. This integration forms the basis for evaluating agents across three core performance metrics:

\begin{itemize} [topsep=1pt,itemsep=0ex,parsep=1ex, leftmargin=*]
    \item \textit{Correctness.} We evaluate correctness by comparing the final network state produced by the LLM agent $\hat{s}_{\text{LLM}}$ to the ground-truth state $s^T$, defined as $s_T$ for constructive tasks and $s_0$ for reactive tasks. Correctness is defined as:
    \begin{equation}
        \textsc{Correct}(Q) = \mathbb{I}\left( \hat{s}_{LLM} \equiv s^*_T \right)
    \end{equation}
    Here, $\hat{s}_{\text{LLM}} \equiv s^*_T$ denotes application-specific state equivalence, where equivalence may be syntactic (e.g., graph isomorphism) or functional (e.g., restored connectivity), depending on tasks.

    \item \textit{Safety.} Safety evaluates whether LLM-generated actions satisfy task constraints $\mathcal{C}_Q$, including structural invariants (e.g., no cross-layer violations) and operational guarantees (e.g., no unauthorized changes, no service disruption). For multi-turn execution with states ${s_0, s_1, \ldots, s_T}$, safety is checked at each step:
    \begin{equation}
        \textsc{Safe}_{\text{all}}(Q) = \mathbb{I}\left( \forall t \in [1, T],\; s_t = \mathcal{E}(s_{t-1}, \hat{a}_{t-1}(\hat{\theta}_{t-1})) \;\wedge\; s_t \models \mathcal{C}_Q \right)
    \end{equation}
    This formulation decouples final correctness from per-step safety, enabling fine-grained evaluation of agent behavior throughout the execution.

    \item \textit{Latency.} Latency captures how quickly and compactly an agent completes the task. We track how many commands the agent issues and the end-to-end latency from query issuance to task completion. Efficient agents solve tasks with fewer commands and minimal latency, which is especially critical in time-sensitive scenarios like failure recovery. 
    
    % \kevin{This part is a bit unclear. In results, we use "latency" instead of "efficiency". Also, "efficiency" sometimes can be related to cost. Suggest using latency throughout.)}
\end{itemize}

% These controlled network emulator environments enable formal execution of agent actions, rigorous outcome validation, and systematic tracking of side effects. By dynamically generating queries through randomized sampling in each evaluation round, we further reduce the risk of data contamination and ensure that agents are consistently tested on diverse, unseen tasks.

With this design, \name dynamically generates queries via randomized sampling in each evaluation round, reducing the risk of data contamination and ensuring agents are consistently tested on diverse, unseen tasks. Guidance on extending \name to new applications is in Appendix~\S\ref{appendix:extend}.
\section{Experiments}
\label{sec:eval}

\textbf{Setup.} We create five AI agents using two base LLMs: GPT-4o~\cite{gpt-4o} and QWen2.5-72B~\cite{yang2024qwen2}. Each model is paired with two prompting strategies: Chain-of-Thought (CoT)~\cite{origin_cot} and Few-shot~\cite{origin_fewshot}. We also evaluate a more advanced agent with ReAct~\cite{yao2023react} on GPT-4o.

\textbf{Representative Tasks.}
To demonstrate generality, we implement three network automation tasks:
\begin{itemize}[topsep=1pt,itemsep=0ex,parsep=1ex, leftmargin=*]
    \item \textit{Capacity Planning (\textbf{CP}).}
    Agents are evaluated on structured planning tasks over a realistic datacenter topology, based on Google’s multi-layer abstraction~\cite{malt}. Tasks include estimating bandwidth and modifying device configurations, spanning 12 action types (e.g., \texttt{add}, \texttt{update}, \texttt{rank}). Safety is checked by enforcing structural constraints and ensuring bandwidth meets minimum thresholds. Latency is measured as end-to-end solution time. (Details in \S\ref{appendix:malt})

\item \textit{Routing Misconfiguration (\textbf{Routing}).}
    Agents diagnose and repair dynamic faults (e.g., broken links or invalid forwarding rules) in ~\citet{mininet}. This involves issuing diagnostic commands, interpreting outputs, and applying fixes to restore connectivity. Safety checks whether modifications improve the state without introducing new issues, and latency is measured by the number of iterations to resolution. (Details in \S\ref{appendix:route})

\item \textit{Microservice Policy Deployment (\textbf{K8s}).}
    Agents troubleshoot misconfigured Kubernetes (K8s) network policies in Google’s open-source microservice demo~\cite{google-k8s-demo}, aiming to restore valid inter-service communication. Tasks involve identifying incorrect ports or overly restrictive rules. Safety metric follows the Routing definition, evaluating whether changes are necessary and effective, while latency counts the steps required for resolution. (Details in \S\ref{appendix:k8s})
\end{itemize}

% Key findings form results:

% Difference in the nature of different applications: 
% distribution of error can be quite different. 
% Coverage and sampling strategy.

% A way to describe capability of each agent. Compare agents by type of error, or some other details.
\begin{figure*}[t]
    \centering
    \begin{subfigure}{0.9\textwidth}
    {\includegraphics[width=1.0\textwidth]{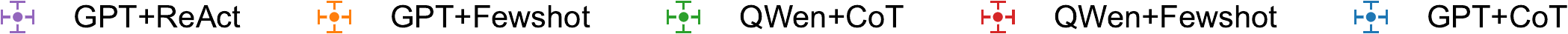}
    }
    \vspace{-3mm}
    \end{subfigure}
    \begin{subfigure}[t]{0.32\textwidth}
    {
        \includegraphics[width=0.99\linewidth]{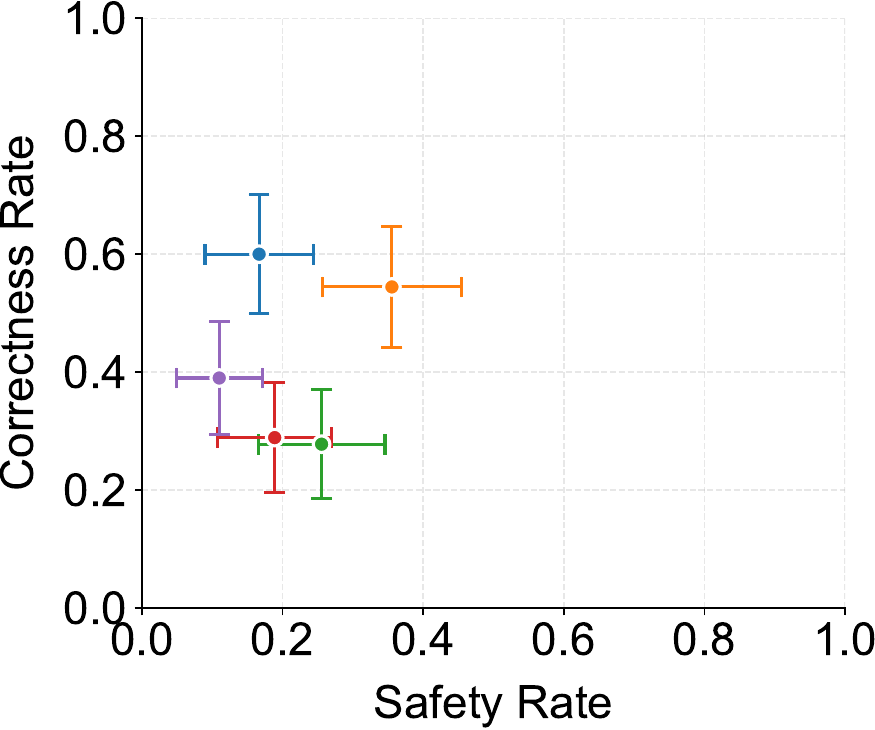}
        \caption{\small CP:100}
        \label{fig:}

    }
    \end{subfigure}
    \begin{subfigure}[t]{0.32\textwidth}
    {
        \includegraphics[width=0.99\linewidth]{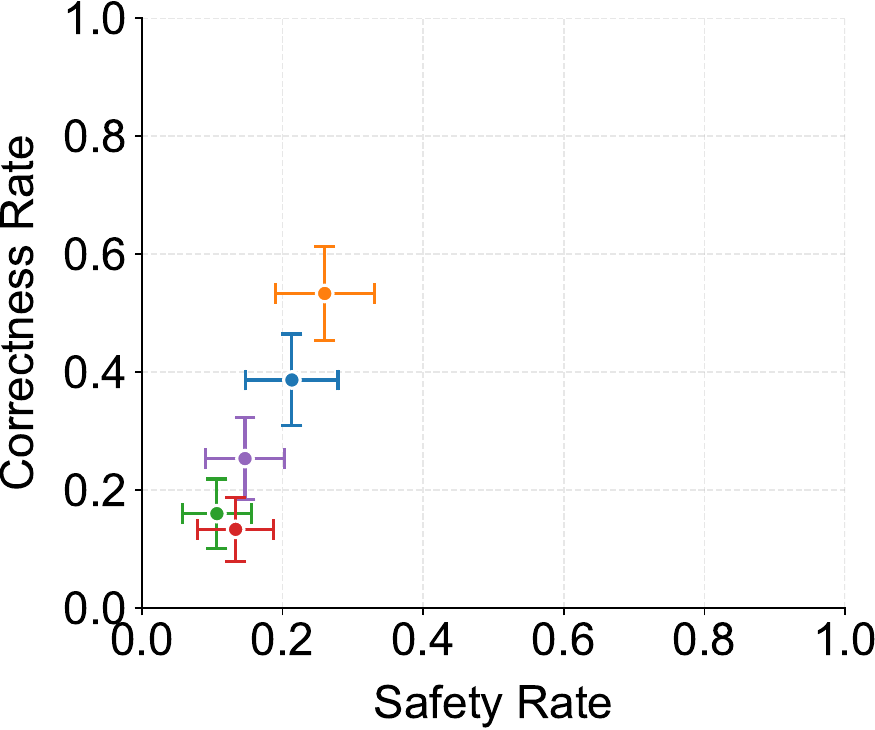}
        \caption{\small Routing:150}
        \label{fig:}

    }
    \end{subfigure}
    \begin{subfigure}[t]{0.32\textwidth}
    {
        \includegraphics[width=0.99\linewidth]{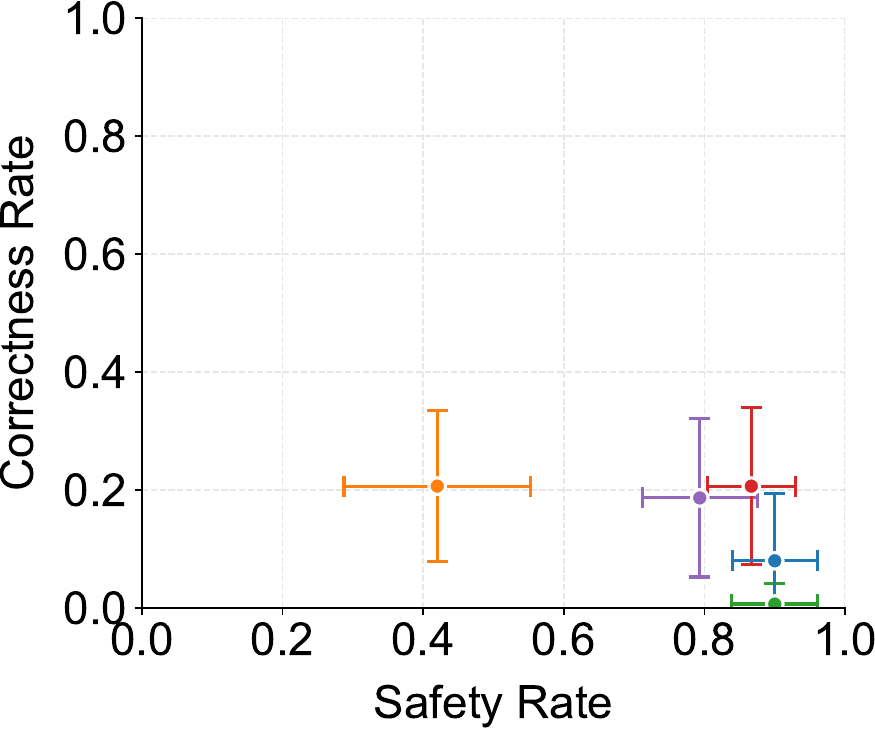}
        \caption{\small K8s:150}
        \label{fig:}

    }
    \end{subfigure}
    
    \begin{subfigure}[t]{0.32\textwidth}
    {
        \includegraphics[width=0.99\linewidth]{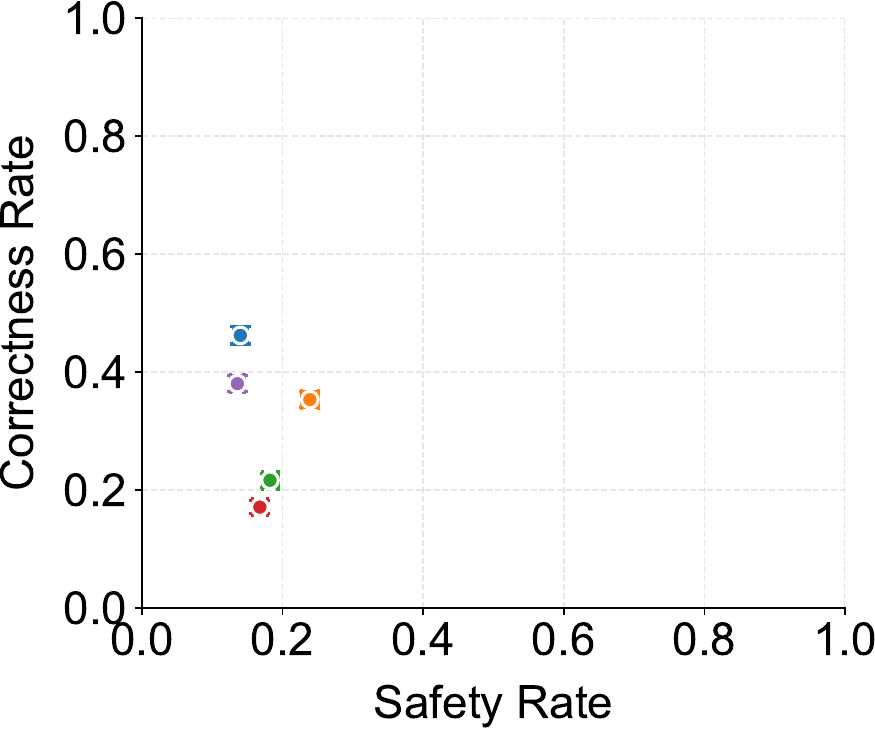}
        \caption{\small CP:5000}
        \label{fig:}

    }
    \end{subfigure}
    \begin{subfigure}[t]{0.32\textwidth}
    {
        \includegraphics[width=0.99\linewidth]{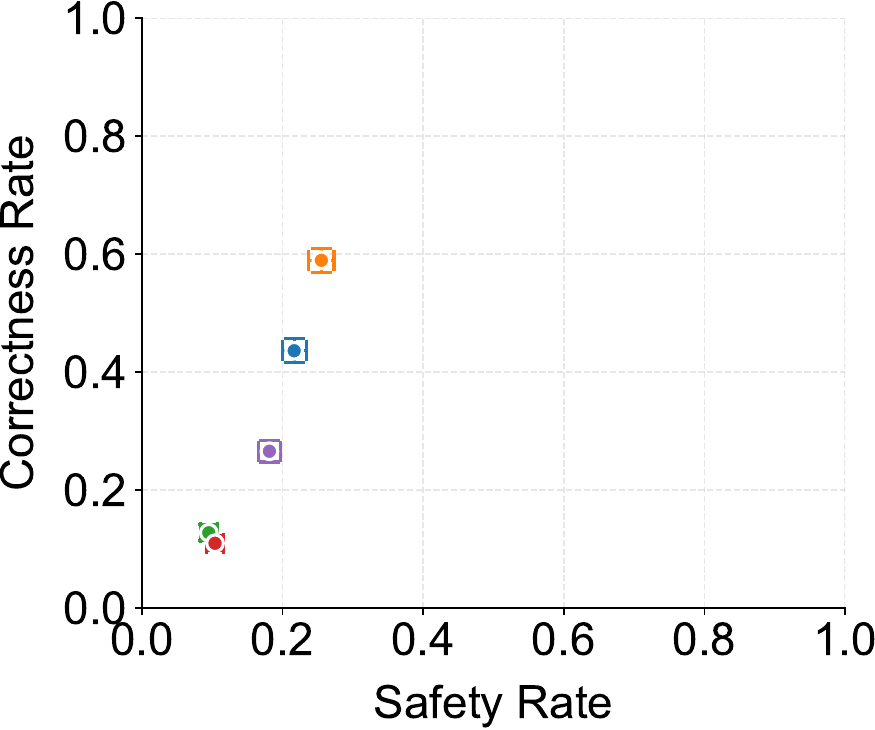}
        \caption{\small Routing:2250}
        \label{fig:}

    }
    \end{subfigure}
    \begin{subfigure}[t]{0.32\textwidth}
    {
        \includegraphics[width=0.99\linewidth]{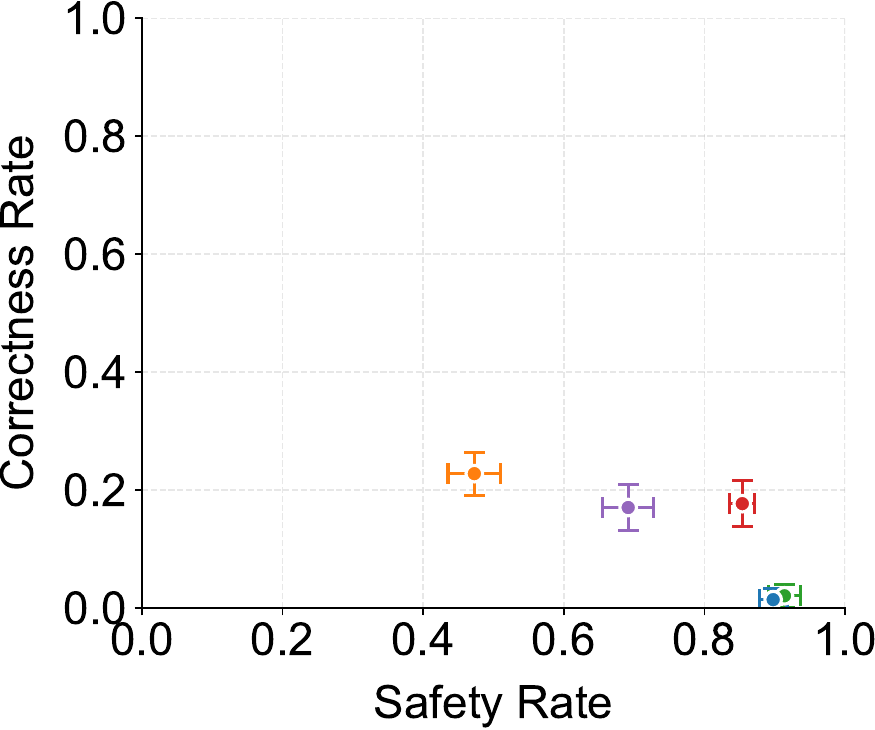}
        \caption{\small K8s:2000}
        \label{fig:}

    }
    \end{subfigure}
    \caption{\name enables dynamic query generation with large query sizes, significantly improving the statistical confidence of agent comparisons.}
    \label{fig:eval:query_size}
    % \vspace{-3mm}
\end{figure*}
\subsection{Reducing Confidence Interval Overlap with Larger Query Size}
\label{eval:confidence_interval}

A key advantage of \name's dynamic generation is its ability to evaluate agents on large, diverse query sets, improving the statistical reliability of comparisons. Since correctness and safety are binary outcomes (i.e., pass/fail per query), we compute confidence intervals using the standard error of the mean (SEM) for a Bernoulli distribution:
$\mathrm{SEM} = \sqrt{\frac{\hat{p}(1 - \hat{p})}{N}},$
where $\hat{p}$ is the empirical success rate and $N$ is the number of queries. We report 95\% confidence intervals as $\hat{p} \pm 1.96 \cdot \mathrm{SEM}$.

As shown in Figure~\ref{fig:eval:query_size}, small query sets (CP:100, Route:150, K8s:150) produce wide error bars and overlapping intervals, making it difficult to distinguish agent performance. For example, on CP:100, GPT+ReAct overlaps by more than 50\% with both QWen+CoT and QWen+Fewshot. Scaling to larger query sets with \name (e.g., CP:5000) eliminates this overlap, revealing GPT+ReAct as the clear winner. Beyond correctness, larger benchmarks also expose agents to richer task variations, reducing overfitting and enabling more robust generalization analysis.

The two-dimensional graph in Figure~\ref{fig:eval:query_size} also highlights the importance of evaluating both correctness and safety, especially when agents have similar correctness rates. For example, on K8s:150, GPT+ReAct and QWen+Fewshot show comparable correctness and safety rates. But on K8s:2000, GPT+ReAct exhibits a noticeably lower safety rate, making it riskier in practice. 
% This shows that large-scale benchmarks with safety metrics are essential for helping network operators choose safer agents for real-world deployment, even when correctness appears similar. 
% \kevin{Draw some observations about correctness vs safety} \lesley{added.}

\subsection{Fine-Grained Evaluation via Complexity-Aware Breakdown}
\label{eval:spider_chart}

\begin{figure*}[t]
\centering
    \begin{subfigure}{0.9\textwidth}
    {\includegraphics[width=1.0\textwidth]{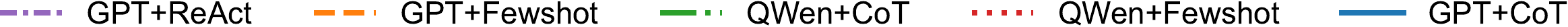}
    }
    \vspace{-2mm}
    \end{subfigure}
    \begin{subfigure}[t]{0.3\textwidth}
    {
        \includegraphics[width=1.0\linewidth]{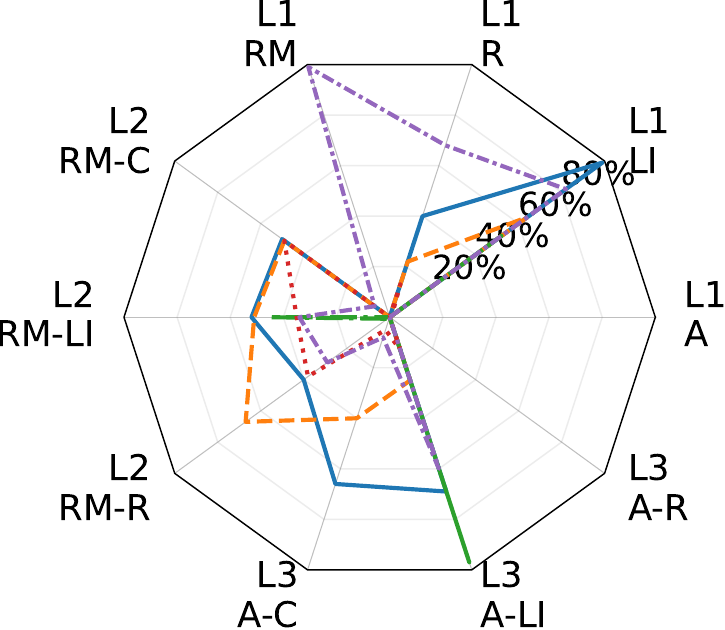}
        \caption{\small CP: Correctness}
        \label{fig:}

    }
    \end{subfigure}
    \begin{subfigure}[t]{0.3\textwidth}
    {
        \includegraphics[width=1.0\linewidth]{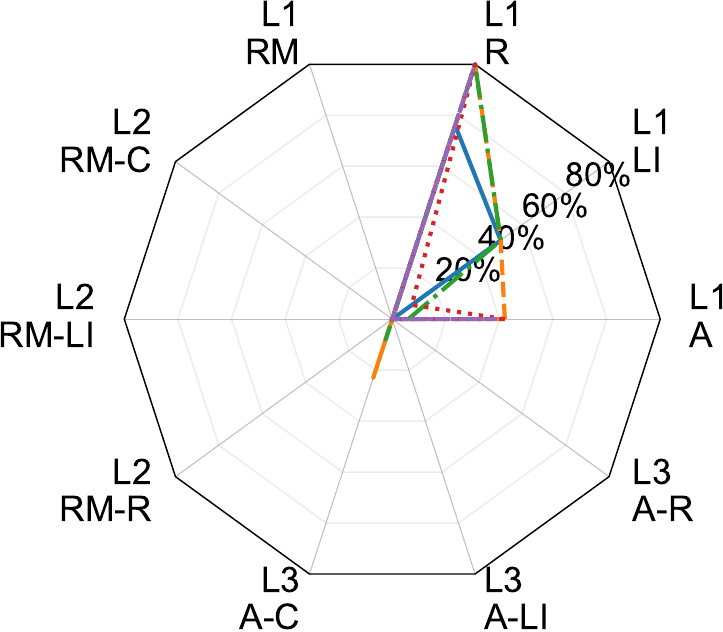}
        \caption{\small CP: Safety}
        \label{fig:}

    }
    \end{subfigure}
    \begin{subfigure}[t]{0.3\textwidth}
    {
        \includegraphics[width=0.99\linewidth]{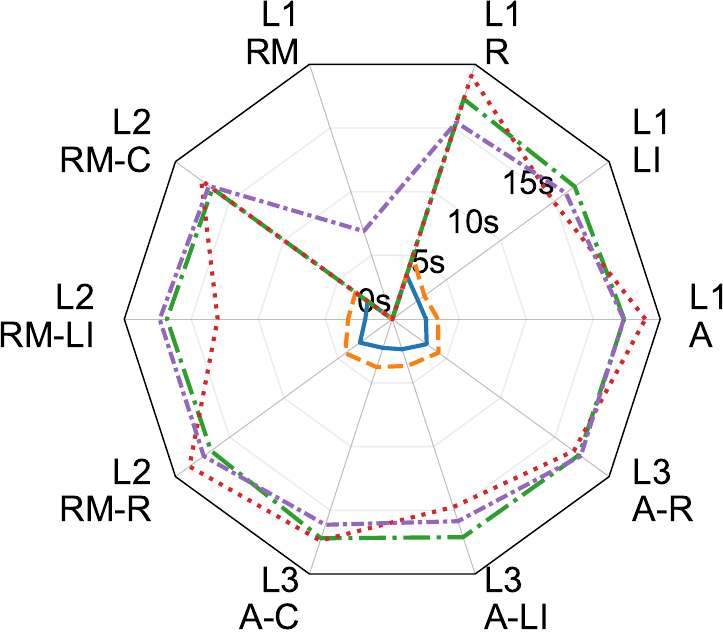}
        \caption{\small CP: Latency}
        \label{fig:}

    }
    \end{subfigure}
    
    \begin{subfigure}[t]{0.32\textwidth}
    {
        \includegraphics[width=0.99\linewidth]{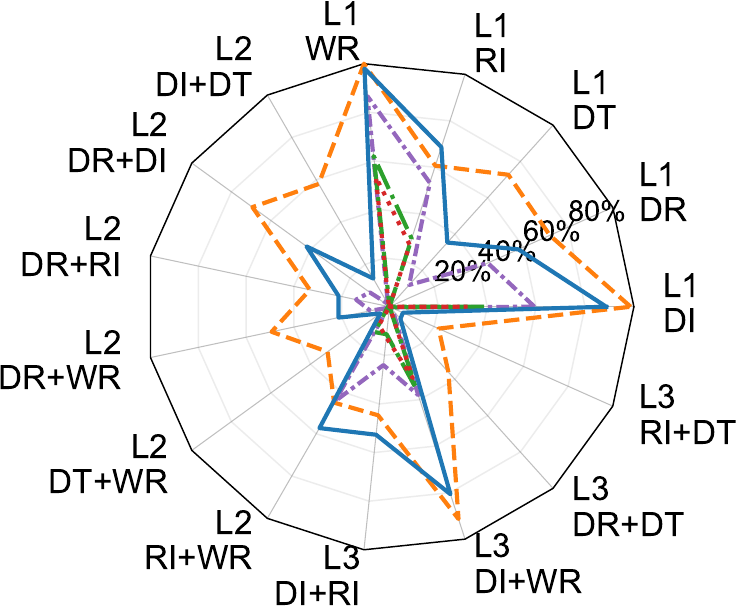}
        \caption{\small Routing: Correctness}
        \label{fig:}

    }
    \end{subfigure}
    \begin{subfigure}[t]{0.32\textwidth}
    {
        \includegraphics[width=0.99\linewidth]{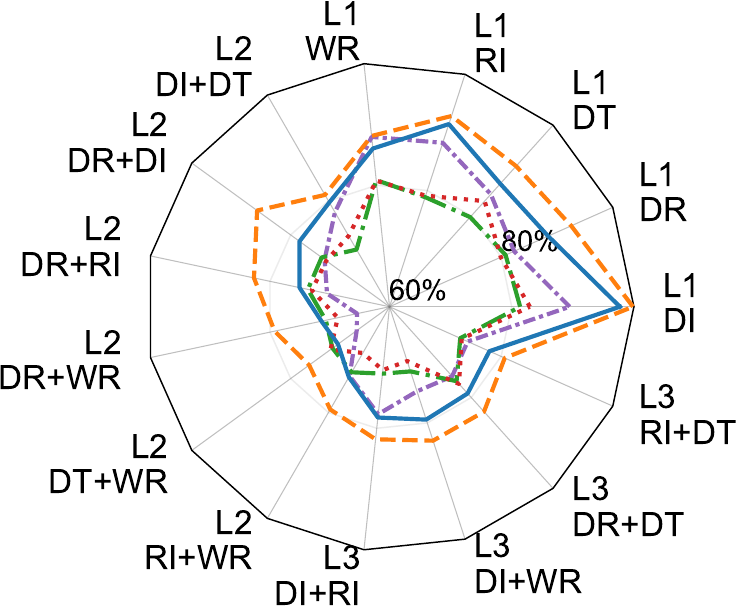}
        \caption{\small Routing: Safety}
        \label{fig:}

    }
    \end{subfigure}
    \begin{subfigure}[t]{0.32\textwidth}
    {
        \includegraphics[width=0.99\linewidth]{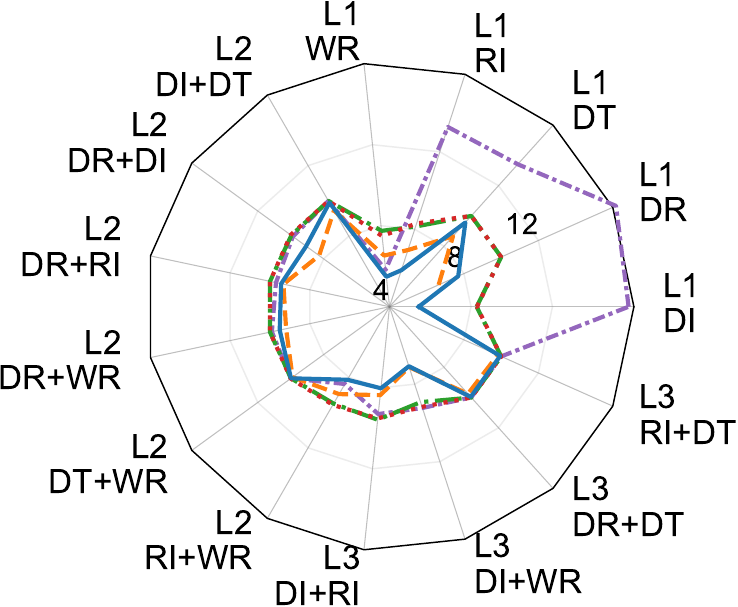}
        \caption{\small Routing: Latency}
        \label{fig:}

    }
    \end{subfigure}
    
    \begin{subfigure}[t]{0.32\textwidth}
    {
        \includegraphics[width=0.99\linewidth]{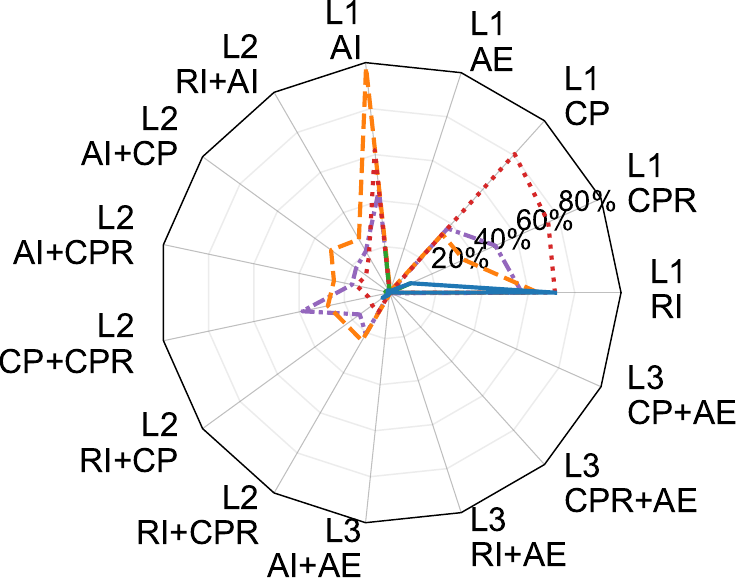}
        \caption{\small K8s: Correctness}
        \label{fig:}

    }
    \end{subfigure}
    \begin{subfigure}[t]{0.32\textwidth}
    {
        \includegraphics[width=0.99\linewidth]{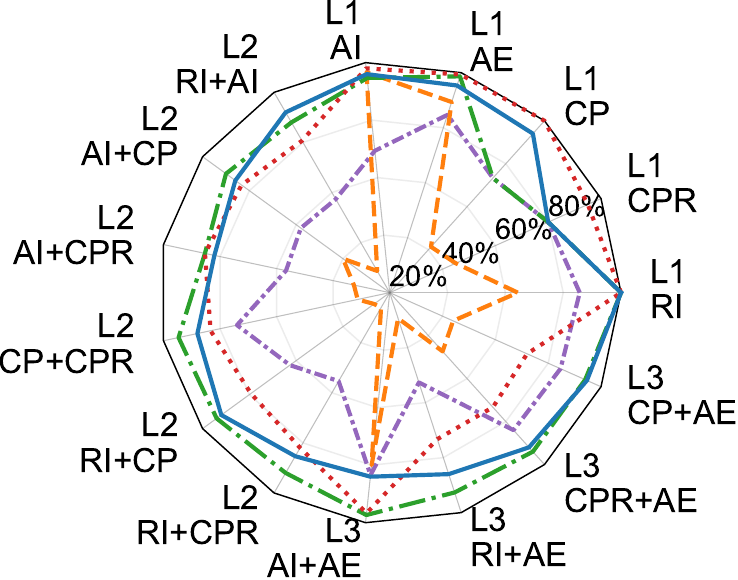}
        \caption{\small K8s: Safety}
        \label{fig:}

    }
    \end{subfigure}
    \begin{subfigure}[t]{0.32\textwidth}
    {
        \includegraphics[width=0.99\linewidth]{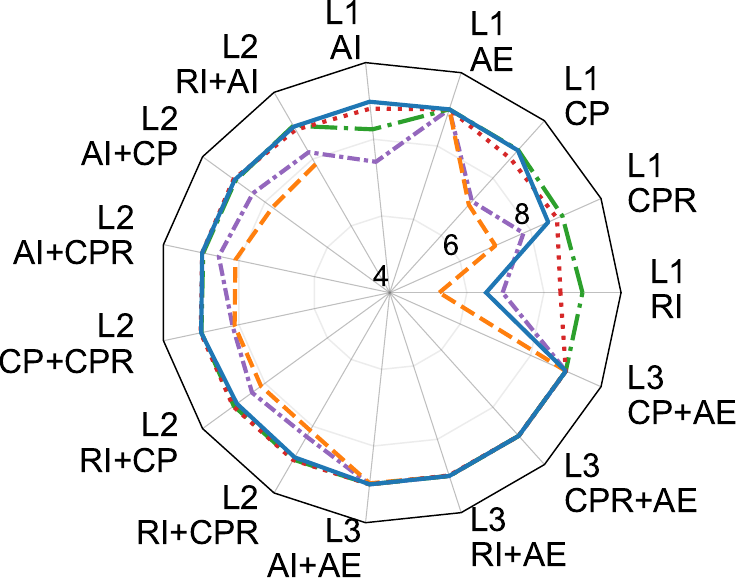}
        \caption{\small K8s: Latency}
        \label{fig:}

    }
    \end{subfigure}
    \caption{Breakdown analysis across complexity levels (L1–L3, gets harder counter-clockwise). Agent strengths vary by application, metric, and task complexity.   
        % \fyy{these spider charts are too difficult to read}
    }
\label{fig:eval:spider}
    \vspace{-5mm}
\end{figure*}
\begin{table}[t]
\centering
\small
\begin{tblr}{
  colspec = {|Q[1.2]|Q[2.5]|Q[2.5]|Q[2.5]|}, % add | for vertical lines
  cell{1}{1} = {c, font=\bfseries},
  cell{1}{2-Z} = {c, font=\bfseries},
  row{1} = {bg=gray9},
  hline{1-2,3,4,5} = {-}{},
  vlines
}
\textbf{Failure Type} &
\textbf{Capacity Planning} &
\textbf{Routing} &
\textbf{Microservice Policy} \\

{Safety\\Violation} &
An LLM adds a switch with 0 connection, violating datacenter constraints. &
An LLM assigns incorrect IPs, breaking existing network connectivity. &
An LLM deletes a running pod, causing immediate service disruption. \\

Control Logic Error &
An LLM calculates capacity at the wrong layer; it did not traverse the topology. &
An LLM misorders commands, assigning an IP before activating the interface. &
An LLM uses incorrect \texttt{patch}, despite examples showing alternatives. \\

Operational Error &
An LLM hallucinates nonexistent node attributes, triggering execution failures. &
An LLM uses incorrect \texttt{systemctl}, trying to edit full system services. &
An LLM omits the namespace in a Kubernetes command, causing it to fail. \\
\end{tblr}
\caption{Example failure types per task. Even with a complete prompt context, LLMs frequently produce incorrect outputs.}
\label{tab:eval:failure}
% \vspace{-6mm}
\end{table}

While aggregated metrics are useful, they often hide critical weaknesses. A model may appear strong on simple queries but fail on compositional tasks or violate safety constraints in complex scenarios. To surface these issues, \name applies complexity control during benchmark generation, annotating each query with action types and difficulty levels (Tables~\ref{tab:append:query_details:malt}, \ref{tab:append:query_details:k8s}, \ref{tab:append:query_details:route} in the Appendix), enabling more fine-grained analysis of correctness, safety, and latency.

As shown in Figure~\ref{fig:eval:spider}, this breakdown reveals clear differences across applications. In datacenter capacity planning (CP), both correctness and safety drop sharply as complexity grows, particularly for GPT-4o+Few-shot. 
% This reflects higher variance in complex tasks, where retrieval-style pattern matching fails to generalize and autonomous reasoning is required. 
This suggests higher variance in complex tasks, where retrieval-style pattern matching fails to generalize and agents must rely on autonomous reasoning.
Across all agents, \texttt{add} operations remain consistently hard (e.g., adding a new device to the datacenter while satisfying capacity-balancing constraints), because they require strict structural constraint satisfaction when introducing new nodes.

In K8s tasks, GPT-4o+Few-shot shows stronger performance in multi-turn, stateful diagnosis. Safety analysis, however, uncovers divergent behaviors. GPT-4o agents are often too aggressive, issuing unsafe fixes such as ``removing ingress+change protocol (RI+CPR)'' in the K8s environment. Other models lean too conservative, frequently failing to act even when safe resolutions exist. Latency patterns vary as well: some agents resolve simple issues efficiently, while others generate long, redundant command sequences even for basic ingress fixes. 

% \kevin{this statement seems to be only true for K8s, not for rounting}

By moving beyond single-number scores, \name provides a deeper view of each agent’s planning strategies, failure modes, and generalization boundaries. 
This analysis strengthens benchmark reliability and provides actionable insights for improving LLM deployment readiness. 
Table \ref{tab:eval:failure} gives additional failure examples of agents in each task.

\subsection{Evaluating Agents Generalization via Supervised Fine-Tuning}
\label{eval:sft_bar}

For constructive tasks with automatically derivable intermediate solutions, \name enables large-scale labeled data generation for supervised fine-tuning (SFT). In these tasks, such as datacenter capacity planning, each query defines a transformation from an initial state $S_0$ to a target state $S_n$. \name then automatically synthesizes the full action sequence $a_1, a_2, \ldots, a_n$ that reaches $S_n$. We use these dynamically generated action traces as ground-truth supervision for SFT.

\begin{figure*}[t]
    \centering
    \begin{subfigure}[t]{0.49\textwidth}
    {
        \includegraphics[width=0.99\linewidth]{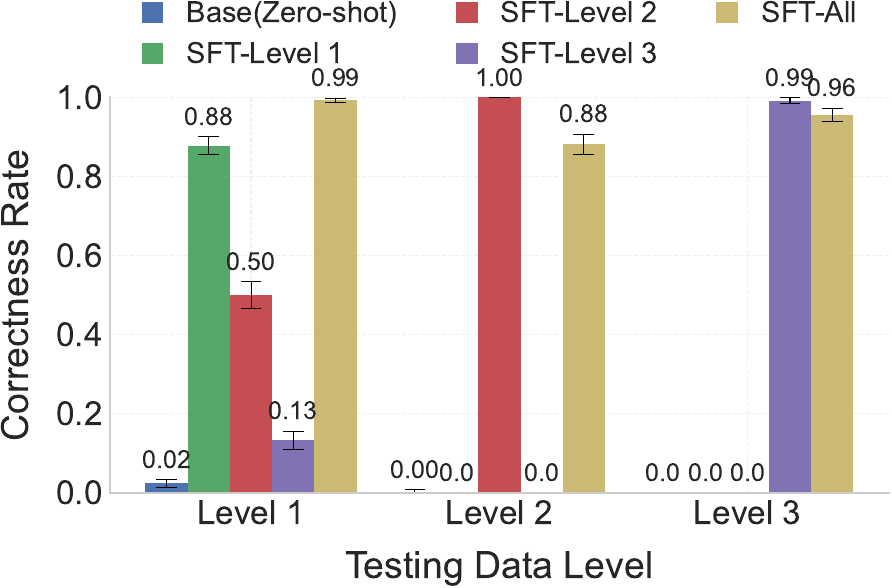}
        \caption{\small Correctness generalization}
        \label{fig:}

    }
    \end{subfigure}
    \begin{subfigure}[t]{0.49\textwidth}
    {
        \includegraphics[width=0.99\linewidth]{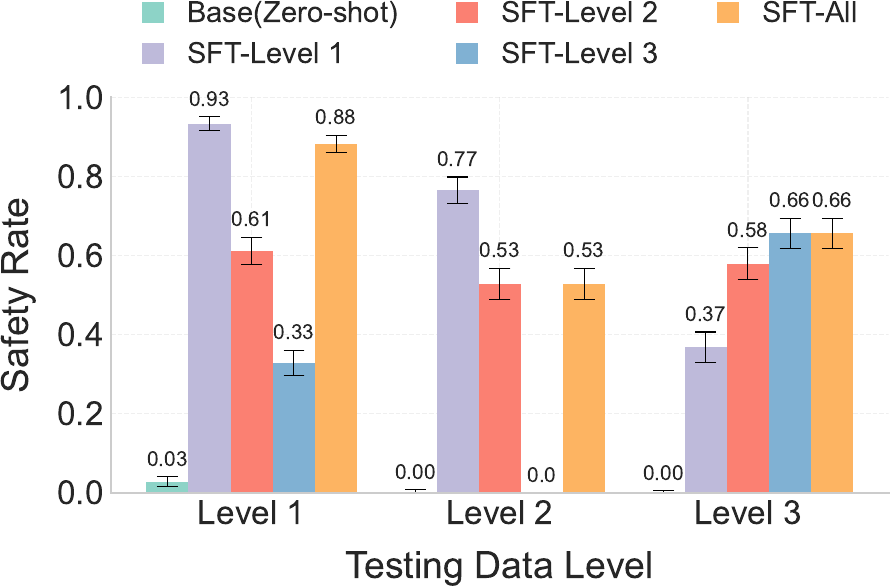}
        \caption{\small Safety generalization}
        \label{fig:}

    }
    \end{subfigure}
    \caption{Supervised Fine-tuned (SFT) model performance at different complexity levels.}
    \label{fig:eval:sft}
    \vspace{-3mm}
\end{figure*}

To study how well a model fine-tuned on one level generalizes to other levels, we fine-tune four Qwen-7B models on different subsets of datacenter capacity planning queries: Level-1 (800), Level-2 (600), Level-3 (600), and a mixed dataset spanning all levels (2000). Each model is then evaluated across all levels using the correctness and safety metrics (Figure~\ref{fig:eval:sft}).

Correctness results show clear overfitting: each model performs well on its own training level but degrades sharply on others. For example, the SFT-Level-2 model reaches perfect correctness on Level-2 but fails on Level-3. Only the mixed-level model generalizes well, maintaining over 0.96 correctness across all levels.

Interestingly, 
safety scores remain more stable. 
Even when correctness drops, models often preserve structural validity, which suggests that safety constraints transfer more readily across task complexities. For example, the SFT-Level-2 model fails on Level-3 in correctness but still achieves reasonable safety, indicating partial transfer of constraint adherence.

Overall, \name provides a practical framework for generating scalable SFT data for constructive network tasks, and this supervision clearly improves model performance. However, we still need rigorous evaluation on unseen task types and configurations to avoid overestimating generalization.
% \kevin{may need to rewrite this paragraph. The conclusion should be for some applications, \name provides a way to generate scalable SFT examples that could improve model performance on these tasks more easily, with some caution about generalizability at unseen levels/combinations/prompts. } \lesley{rewrote this one, plz check}

\section{Conclusion and Use Cases of \name}
\label{sec:use_case}
We present \name, a unified pipeline for dynamic LLM evaluation in real-world network applications. By abstracting tasks into \textit{state–action} form, \name enables controllable query generation with automatic ground-truth derivation. Its integration with high-fidelity emulators supports execution-time validation across correctness, safety, and latency. As such, \name offers a practical foundation for developing, evaluating, and debugging AI agents in safety-critical network domains. \name also enables two prominent future use cases below.

\subsection{Use Case 1: Enabling Post-Training RL in \name's Environments}
\label{sec:discussion:fine_tune_RL}

Many reactive tasks lack step-level ground truth, making supervised fine-tuning infeasible and positioning RL as a natural alternative. However, RL requires reliable environments that \textbf{support interactive execution and feedback for reward computation}. \name addresses this gap by integrating emulators that automatically generate step-wise feedback in response to agent actions, enabling structured RL training and evaluation.

To test this, we fine-tune a QWen2.5-0.5B model (the largest feasible model we can fine-tune with our GPUs) using GRPO from TRL~\cite{trl} in a Mininet routing environment. Rewards include –100 for invalid commands, +10 for valid diagnostics, and +100 for correct fixes. 
% The RL-finetuned model does not fully solve routing issues but consistently produces valid Mininet commands, outperforming its zero-shot baseline. This shows RL can shape useful policies even at small scales, though model capacity itself may limit its improvement in this experiment. 

\begin{figure*}[t]
    \centering
    \begin{subfigure}[t]{0.49\textwidth}
    {
        \includegraphics[width=0.99\linewidth]{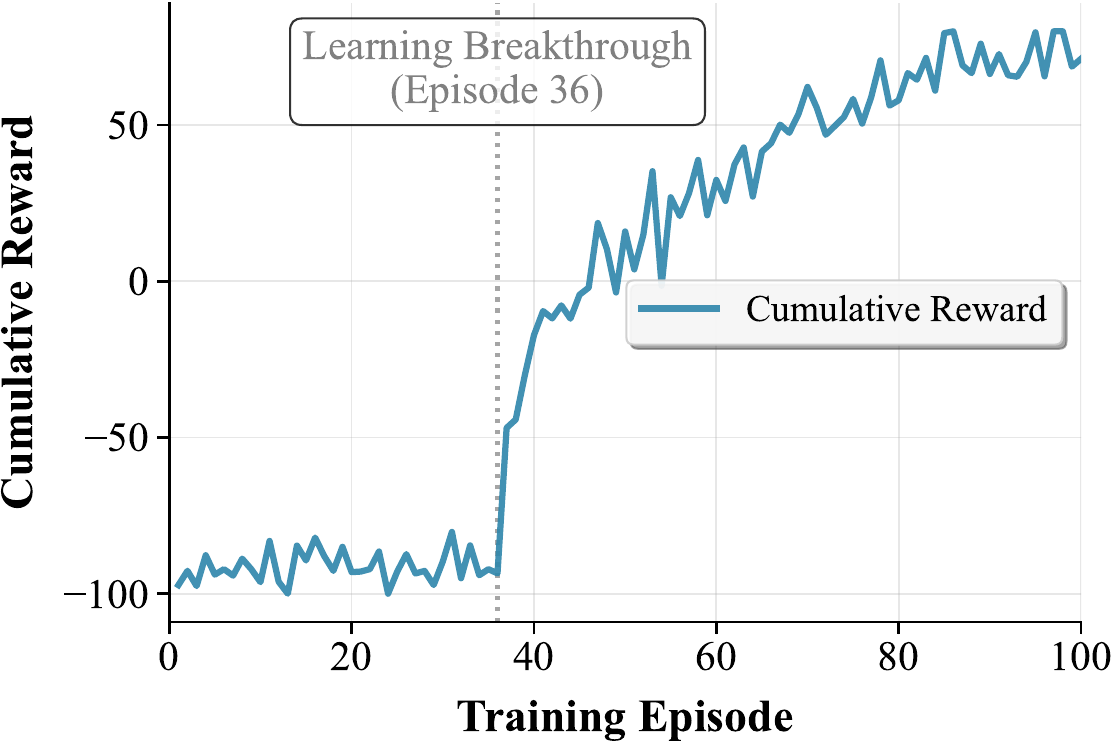}
        \caption{Learning curves under GRPO}
        \label{fig:rl:curve}

    }
    \end{subfigure}
    \hspace{0.03\textwidth} 
    \begin{subfigure}[t]{0.45\textwidth}
    {
        \includegraphics[width=0.99\linewidth]{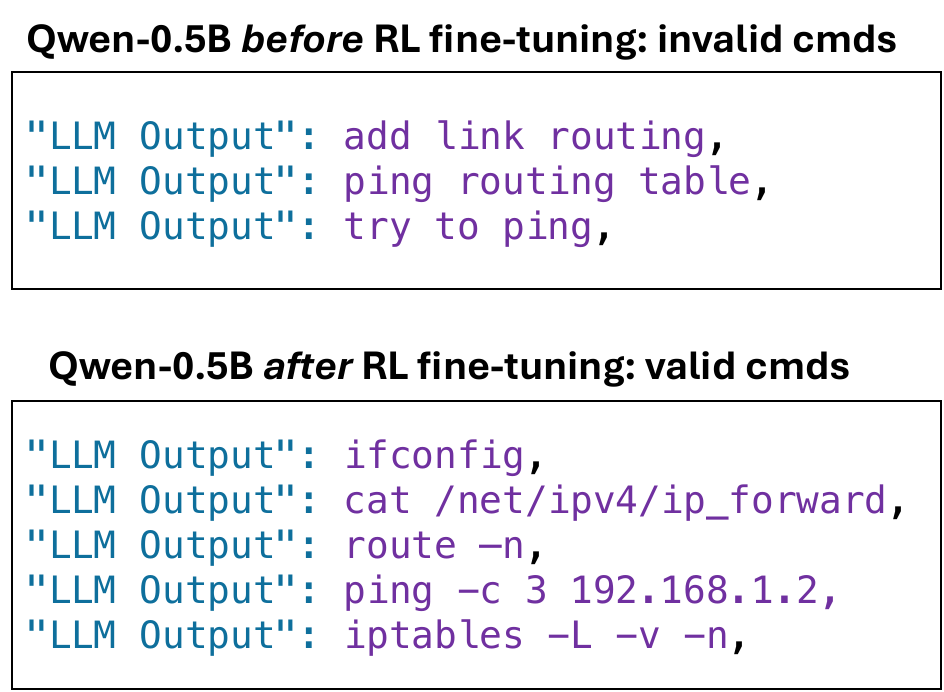}\caption{Qualitative improvements after RL tuning.}
        \label{fig:rl:example}
    }
    \end{subfigure}
    \caption{Preliminary RL fine-tuning analysis with \name.}
    \label{fig:rl}
\end{figure*}

Figure \ref{fig:rl:curve} shows the RL training curves. In the first 36 episodes, the training curve is noisy as the agent is still exploring. After this point, the cumulative reward begins to rise steadily, indicating that the agent is learning from environment feedback.
Figure \ref{fig:rl:example} shows a qualitative before/after comparison. Before RL fine-tuning, with Qwen-0.5B (CoT) keep outputting some random commands that are not valid for the Mininet emulator (e.g., try to ping). After fine-tuning, it generates valid diagnostic and modification commands (e.g., ping -c 3 192.168.1.2). 
While overall correctness remains limited, these results imply that the environment provides learnable feedback.

This experiment shows initial feasibility that \name can be used as a RL training environment.
Beyond post-training RL, \name could potentially enable closed-loop self-improvement: as agents refine their reasoning traces and generate higher-quality action sequences, those traces can be incorporated into future RL episodes to improve agents. 

\subsection{Use Case 2: Probing Agents with Adversarial Examples}
\label{sec:discussion:adversarial}
Understanding agent failure modes is critical for deployment. Unlike static benchmarks that cover a narrow task slice, \name can dynamically generate adversarial test cases targeting specific weaknesses. By analyzing error types, failure modes, and inconsistent reasoning traces, we can identify task configurations that consistently degrade performance.

For instance, \name exposes fine-grained control knobs (e.g., topology size, failure types, task complexity) that can be tuned to explore the model's capability boundaries. We propose RL or heuristic-guided sampling to iteratively generate harder queries based on prior failures. Over time, this adversarial loop reveals critical limitations, giving developers concrete insights into where generalization breaks and where extra training or safeguards are required.

\subsection{Discussions on Generalization from Emulators to Real Deployments.}
A key question for any benchmark built on emulators is how well performance transfers to production environments. NetArena does not claim that success in an emulator directly implies deployment readiness. Real systems introduce additional sources of complexity, including noisy observability, missing instrumentation, operational drift, heterogeneous tooling, and hidden dependencies that are difficult to fully reproduce in a controlled environment. As a result, emulator performance should be interpreted as a measure of an agent's ability to reason over structured operational tasks under realistic, but still simplified, conditions.

That said, emulators remain valuable because they preserve the core elements that drive many operator workflows: stateful system evolution, executable actions, delayed feedback, and safety-critical side effects. In our design, we focus on faithfully capturing these causal and structural properties rather than reproducing every low-level deployment detail. This makes NetArena useful for stress-testing planning, diagnosis, and remediation behavior before real-world trials. 
Narrowing the sim-to-real gap further for each application remains important future work.

\noindent \textbf{Acknowledgments}. This work was supported in part by the U.S. NSF grants SaTC-2415754 and CNS-2415758.
We thank the anonymous reviewers for the constructive feedback.

% \subsection{Generalize the benchmark generation framework to other domains like Robotics}

% \subsection{Safety-Centric Evaluation for Network Environments.}
% In networked systems, correctness alone is not enough—LLM-generated actions must also satisfy strict safety constraints to prevent service disruption, policy violations, or security risks. Our benchmark framework is designed to foreground these concerns by embedding domain-specific safety checks into each evaluation. For instance, we verify that datacenter planning actions preserve topology hierarchies, routing fixes do not create forwarding loops, and Kubernetes policy updates do not expose sensitive services. These structured safety conditions are enforced through executable validators that run in simulators and emulators, enabling automated and repeatable detection of unsafe agent behavior.

% To go beyond fixed safety tests, our benchmark dynamically generates a diverse range of edge cases that stress different failure modes. By systematically perturbing initial states and query goals—such as misaligned policy rules or conflicting route updates—we expand the coverage of safety-critical scenarios far beyond what static datasets allow. Moreover, by tracking which failure types frequently occur across agents, we can highlight recurring unsafe behaviors and guide targeted intervention, whether through prompt tuning, reward shaping, or policy constraints. This continuous expansion of safety coverage is essential for building trustworthy LLM agents in real-world network operations. 

\newpage

\bibliography{iclr2026_conference}
\bibliographystyle{iclr2026_conference}

\appendix
\newpage
\appendix
\onecolumn

% \section{Details of Applications}
% \section{Appendix}

% \section{Compute Resources for Experiments}
% \label{sec:appendix:compute}

% The experiments are conducted on a compute system equipped with dual-socket AMD EPYC 7V13 64-core processors, totaling 96 threads. The system includes 4× NVIDIA A100 GPUs, each with 80GB of memory, providing 320GB total GPU memory and full support for CUDA 12.8. It has 866 GB of system RAM with negligible usage at runtime and no swap configured. For storage, the machine is provisioned with multiple disks: two 1TB SATA drives, a 256GB disk, and four 894GB NVMe SSDs, including one mounted at /datadisk. This setup supports large-scale evaluation workloads with ample compute, memory, and I/O capacity.

\section{Extend \name to New Applications}
\label{appendix:extend}
\name is designed for easy extensibility across diverse network and system domains through a standardized API. To add a new application, users define three key components: (1) the application's \textit{state} and \textit{action} spaces, (2) a backend emulator or simulator that connects to \name's evaluation framework (such as Mininet or Kubernetes), and (3) three core metrics customized to the task. Table~\ref{tab:steps:extend} presents this process in a clear step-by-step format. By lowering the barrier to incorporating realistic, execution-grounded tasks, \name becomes a generalizable platform for benchmarking LLM agents in real-world infrastructure reasoning scenarios. We encourage the community to extend \name to new protocols, topologies, and deployment settings, supporting robust and reproducible agent evaluation at scale.
\begin{table}[htb]
\centering
\small
\begin{tblr}{
  colspec = {|Q[0.2,c]|Q[2,l]|X[6,l]|}, % three columns: number, step, description
  cell{1}{1} = {c, font=\bfseries},
  cell{1}{2} = {c, font=\bfseries},
  cell{1}{3} = {c, font=\bfseries},
  row{1} = {bg=gray9},
  hline{1-8} = {-}{}, % horizontal line for every row
  vlines,
}
No. & Step & Description \\

1 & Emulator Selection &
Select a high-fidelity emulator that matches the target network application. 
Examples: \texttt{Kubernetes} (microservices), \texttt{Mininet} (routing/switching). \\

2 & Topology Generator &
Implement a topology creation function. 
Input parameters: number of nodes, edge density, topology constraints (e.g., connectivity). 
Output: a full topology object deployable in the emulator. \\

3 & Task Type Selection &
Choose the task category: Constructive tasks -- LLMs propose configurations; Reactive tasks -- LLMs respond to injected errors or alerts. \\

4 & Operation Library (Constructive) &
Define basic operations representing atomic configuration steps. These should be modular and composable, and parameterizable (e.g., \texttt{AddNode}, \texttt{LinkSwitch}). Used to generate complex configuration queries. \\

5 & Error Injection (Reactive) &
Define typical error types for reactive diagnosis tasks (e.g., link failure, misconfiguration). These errors are injected into the environment to test LLM diagnostic capabilities. \\

6 & Evaluation Metrics &
Define evaluation criteria: Correctness -- does the LLM achieve the intended effect?; Safety -- does it interrupt the existing service?; Latency -- how fast does the LLM generate valid responses? \\
\end{tblr}
\caption{\rev{Construction steps for creating new applications in \name.}}
\label{tab:steps:extend}
\end{table}

\section{Details for Datacenter Capacity Planning }
\label{appendix:malt}
% \begin{item
%     \item Background of the application, why is this application important and useful in real world.  (Introduce to ML people what it is about)
%     \item How is LLM used in this application. (What do ppl ask LLM to do)
%     \item Environment details: Topology creation; Simulator usage; Give an example on the simulator feedback. Metric definition (correctness, safety, latency, etc)
%     \item Benchmark creation: How to create query? Basic actions/errors? How to define complexity; How to verify results (create ground truth for dynamic queries)?
%     \item Results: How many LLMs? Breakdown results on each LLM  
% \end{itemize}

The first application focuses on applying LLM agents for datacenter capacity planning—a critical aspect of network lifecycle management that optimizes resource utilization and minimizes provisioning costs. Effective capacity planning depends on accurate network topology representations at different abstraction levels. High-level abstractions help network operators evaluate overall bandwidth needs between data centers, while detailed low-level views enable engineers to manage individual device configurations and connections efficiently.

Based on Google's multi-layer topology abstraction~\cite{malt} and publicly available datasets, we build a simulation environment modeling a realistic datacenter topology with 5,493 nodes and 6,424 edges, encompassing 10 distinct device types such as packet switches, ports, and chassis. Nodes have attributes like physical port capacity and are interconnected following hierarchical constraints reflective of real datacenter structures.

To enable dynamic interaction between LLM agents and representative capacity-planning tasks, we define 12 operational types (e.g., `update,' `add,' `count,' `rank'), each supporting a wide range of diverse and detailed queries. For instance, an `add' operation may involve simple tasks, like attaching a new port to a switch, or complex scenarios, such as integrating a new packet switch into an aggregation block. Each query instance is generated dynamically and randomly via i.i.d. sampling, reducing data contamination risk.
LLM agents generate executable code for each query, which is evaluated against dynamically generated ground-truth code execution results. A query is considered successfully resolved if the agent’s generated code executes correctly, and without introducing any security vulnerabilities or violations of the datacenter hierarchical constraints.

\subsection{Application environment}
To enable LLM agents to interact with data center capacity planning, we adopt Google's multi-layer topology abstraction~\cite{malt} and use their released datasets as the foundational topology. We also develop a simulator to evaluate the performance of deployed capacity planning algorithms across key metrics.

\textbf{Topology.} The capacity planning topology consists of 5,493 nodes and 6,424 edges. Each node represents different device types within the data center (e.g., packet switch, port, chassis), with a total of 10 distinct node types. Each node may also have attributes; for instance, a port may have a physical capacity attribute, which specifies its capacity. 
These nodes are interconnected by edges, where, for example, an edge between a packet switch and a port indicates that the switch contains the corresponding port. To simulate the abstraction of a real data center, not all nodes can be arbitrarily connected. Figure \ref{fig:appendix:malt} illustrates the hierarchical dependencies between the nodes.
\begin{figure}[htb]
    \centering  
    \includegraphics[width=0.65\textwidth]{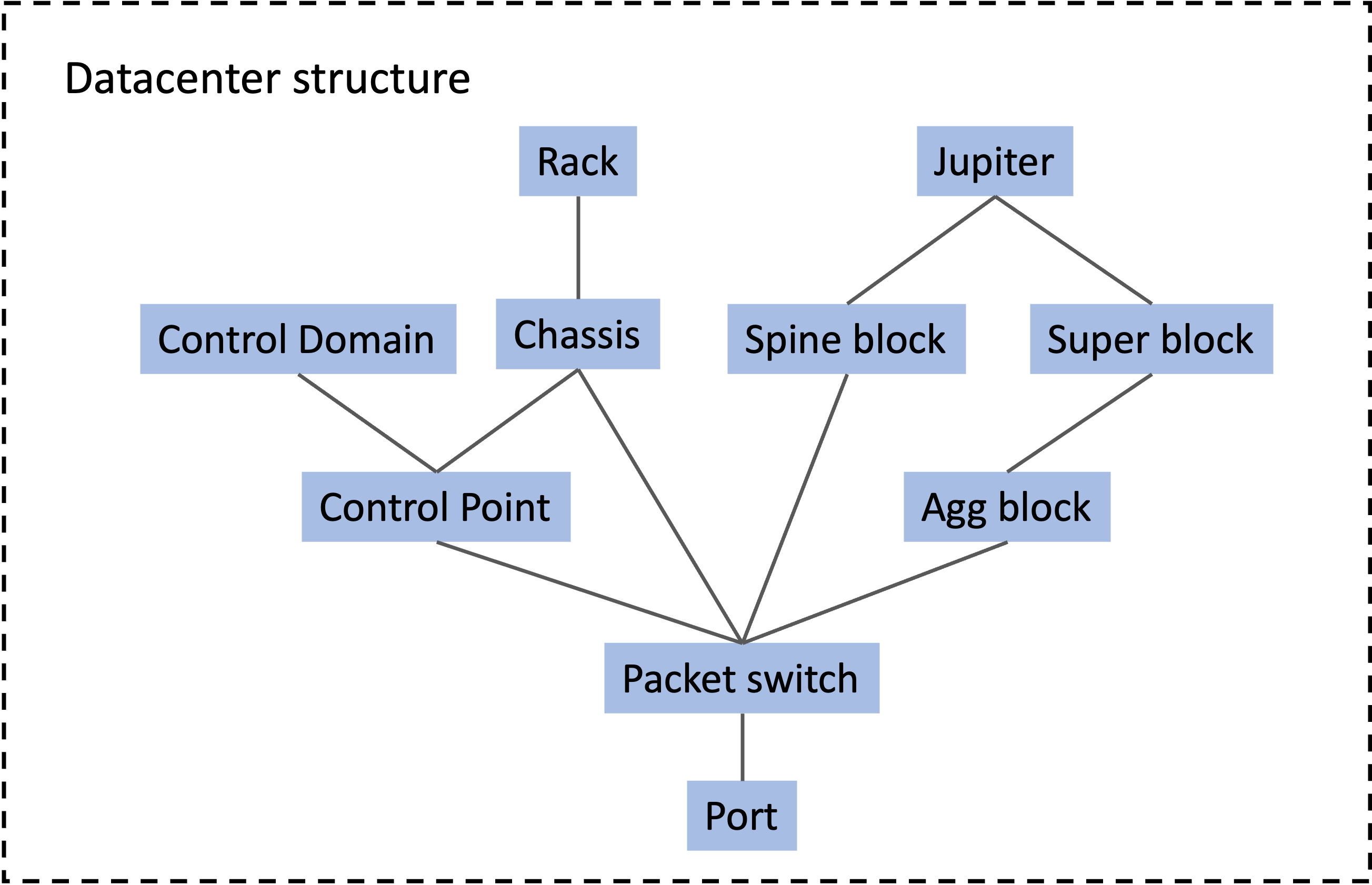} 
    \caption{Hierarchical dependencies between nodes in the datacenter topology.}
    \label{fig:appendix:malt}
\end{figure}

\subsection{Dynamic benchmark creation}

\textbf{Basic Operations and Operands.} 
We implement six fundamental operations: \textit{add}, \textit{count}, \textit{update}, \textit{remove}, \textit{list}, and \textit{rank}. Each operation defines a specific action, while the operands represent the datacenter entities or elements the operations act upon. A user query is interpreted as a combination of these operations applied to dynamically generated operands based on the query type. 
For example, consider the query: \emph{``If we add a new packet switch to each chassis, what will be the new total capacity on all chassis?"} This query can be broken down into the following steps: (1) \textit {List }all chassis nodes. (2) \textit{Add} a new packet switch node to each chassis node. (3) \textit{Count} the total capacity on updated chassis nodes. 

\textbf{Query Complexity Control.}
Using the basic operations, queries can be dynamically constructed by combining them with appropriate operands. To manage complexity, we categorize queries based on two factors: the number of operations involved and the type of control sequence utilized. Control sequences can vary in structure, including sequential combinations of multiple actions, conditional \textit{If-Else} statements, loops such as \textit{For-loops} followed by sequential actions, or \textit{For-loops} combined with \textit{If-Else} statements. 
The operands associated with these basic operations are determined dynamically for each query, allowing the framework to remain flexible and adaptable to diverse query types and scenarios.

\textbf{Ground Truth Generation.} Dynamically generating ground truth algorithms is the most complex process. To accomplish this, we first manually implement the algorithm for each basic operation, supported by dynamic operands, as functional modules. These modules are then combined into templates to generate more detailed queries and their corresponding ground truth. For instance, when the randomly selected operation types are ``add'' and ``count'', the system randomly selects new node types and parent nodes. A unique node name is created, and a corresponding natural language query is formed: ``Add ${child\_node\_name}$ to ${parent\_node\_name}$. Count the ${child\_node}$ in ${parent\_node\_name}$ in the updated graph.'' The system subsequently generates a Python function by combining the ``add'' and ``count'' algorithms into a new function, which represents the ground truth. This function includes the steps of adding the selected node to the graph, linking it to its parent node, and performing a counting query on the updated graph.

\subsection{LLM agents usage}

\textbf{Emulator.} 
To emulate datacenter capacity-planning tasks with large language model (LLM) agents, we build a Python evaluation pipeline that simulates real-world decision-making by executing model-generated code on a dynamic infrastructure graph and checking both correctness and safety. The evaluator initializes with a network graph, an LLM agent, and the corresponding prompting strategy. When a user issues a query, the agent generates Python code to analyze or modify the graph, and the framework executes this code to produce a structured result.

A built-in safety checker then verifies that the updated graph satisfies datacenter constraints, including valid node types, edge formats, topological hierarchy, bandwidth configuration, and switch port presence. In parallel, the framework executes and evaluates a golden (ground-truth) answer in the same way. It then compares the agent output with the ground truth using type-specific criteria, such as graph isomorphism for structural outputs or exact equality for lists and text, and records correctness, safety violations, and execution latency.

The framework also captures errors such as invalid output types, flawed logic, and unsafe graph mutations, and stores them in a structured JSON format for downstream analysis. This emulator enables robust benchmarking of LLM agents under system-level constraints and supports the generation of diverse labeled data for supervised fine-tuning and safer deployment of AI agents in datacenter environments.
\textbf{Prompts.} We provide the initial prompt as below.

\begin{PromptBlock}{Prompt 1: Datacenter Capacity Planning}
\label{prompt:malt}
Generate the Python code needed to process the network graph to answer the user question or request. The network graph data is stored as a \pcode{networkx} graph object. The Python code you generate should be in the form of a function named \pcode{process\_graph} that takes a single input argument \pcode{graph\_data} and returns a single object \pcode{return\_object}. The input argument \pcode{graph\_data} will be a \pcode{networkx} graph object with nodes and edges.

Graph Structure:
\begin{itemize}[leftmargin=*, itemsep=2pt]
    \item The graph is directed, and each node has a \pcode{name} attribute to represent itself.
    \item Each node has a \pcode{type} attribute, in the format of \pcode{EK\_TYPE}. Each node can have other attributes depending on its type.
    \item Each directed edge also has a \pcode{type} attribute, which can include \pcode{RK\_CONTAINS} or \pcode{RK\_CONTROL}.
    \item You should check relationships based on edges and check names based on node attributes.
\end{itemize}

\plabel{Node Type Hierarchy:}
\begin{itemize}[leftmargin=*, itemsep=2pt]
    \item \pcode{EK\_JUPITER} contains \pcode{EK\_SPINE\_BLOCK}
    \item \pcode{EK\_SPINE\_BLOCK} contains \pcode{EK\_AGG\_BLOCK}
    \item \pcode{EK\_AGG\_BLOCK} contains \pcode{EK\_PACKET\_SWITCH}
    \item \pcode{EK\_CHASSIS} contains \pcode{EK\_CONTROL\_POINT}
    \item \pcode{EK\_CONTROL\_POINT} contains \pcode{EK\_PACKET\_SWITCH}
    \item \pcode{EK\_RACK} contains \pcode{EK\_CHASSIS}
    \item \pcode{EK\_PACKET\_SWITCH} contains \pcode{EK\_PORT}
    \item \pcode{EK\_SPINE\_BLOCK} contains \pcode{EK\_PACKET\_SWITCH}
    \item \pcode{EK\_CONTROL\_DOMAIN} contains \pcode{EK\_CONTROL\_POINT}
    \item \pcode{EK\_CHASSIS} contains \pcode{EK\_PACKET\_SWITCH}
    \item \pcode{EK\_JUPITER} contains \pcode{EK\_SUPER\_BLOCK}
    \item \pcode{EK\_SUPER\_BLOCK} contains \pcode{EK\_AGG\_BLOCK}
\end{itemize}

\pimp{Adding Nodes:}
\begin{itemize}[leftmargin=*, itemsep=2pt]
    \item Adding new nodes requires considering the attributes of the new node.
    \item You should also add edges based on their relationships with existing nodes.
    \item The name to add on each layer can be inferred from the new node's name string.
\end{itemize}

\pimp{Capacity Calculation:}
\begin{itemize}[leftmargin=*, itemsep=2pt]
    \item Packet switch nodes have a switch location attribute \pcode{switch\_loc}.
    \item \pcode{PORT} nodes have an attribute \pcode{physical\_capacity\_bps}.
    \item When calculating the capacity of a node, sum the \pcode{physical\_capacity\_bps} on the \pcode{PORT} nodes within the hierarchy that contains this node.
\end{itemize}
\end{PromptBlock}

\subsection{Evaluation metrics}
LLM agents are tasked with queries related to generating Python-based capacity planning algorithms. These algorithms take the original network graph as input and produce outputs that depend on the query, either direct answers or updated topologies.
The simulator environment evaluates the following metrics. 
\begin{itemize}[topsep=1pt,itemsep=0ex,parsep=1ex, leftmargin=*]
    \item \textbf{Correctness:} The output's correctness is evaluated by comparing the LLM-generated results with the ground truth. If the outputs match exactly, the LLM's answer is labeled as correct.
    \item \textbf{Safety:} Safety ensures the structural and attribute integrity of the network graph by verifying that the LLM's output adheres to all defined constraints. This includes validating node types, edge types, and hierarchical relationships, checking the presence of mandatory attributes (e.g., physical capacity for PORT nodes), ensuring no isolated nodes exist, and enforcing connectivity rules between related components.
    \item \textbf{Latency:} Latency measures the execution time of the Python code generated by the LLM. This metric excludes the time spent on LLM prompting and response, focusing solely on the effectiveness and runtime performance of the generated solution.
\end{itemize}

\section{Details for Routing Misconfigurations}
\label{appendix:route}
The second application focuses on applying LLM agents for network routing configuration troubleshooting—an essential network management task that involves quickly identifying and resolving routing-related issues such as misconfigured paths, link failures, and network congestion. Effective troubleshooting is critical in practice, as unresolved routing problems can cause severe performance degradation or outages. For instance, a major outage at Meta in October 2021 was caused by faulty configuration changes to backbone routers. This misconfiguration disrupted communication between data centers, leading to a global service outage and significant downtime \cite{meta-outage}.

To evaluate LLM agents on routing troubleshooting tasks, we construct a simulated network environment using Mininet, featuring a router connected to multiple switches and hosts organized into distinct subnets, enabling realistic network interactions. Within this environment, we dynamically inject various routing misconfigurations—such as incorrect forwarding rules or broken links—that disrupt connectivity and cause failures in the pingmesh (host-to-host connectivity) tests. LLM agents is tasked to perform diagnostics by executing network commands (e.g., inspecting routing tables, interface statuses, and error logs), analyzing the results, and proposing corrective actions. An evaluation query is considered successfully resolved if, after the agent's intervention, the network connectivity is restored, verified by the successful execution of the `pingall()' method, without introducing new security issues.

\subsection{Application environment}
To evaluate the network troubleshooting capabilities of LLMs, we designed an experimental setup where LLM behaves like a network engineer to interact with a simulated network environment. Among the different network simulation tools available, we selected Mininet for its lightweight nature, ease of use, and support for custom dynamic topologies. Within Mininet, we created dynamic network topologies and intentionally injected errors to simulate realistic network faults. The LLM was then tasked with diagnosing and resolving these issues, demonstrating its ability to perform automated troubleshooting in a controlled environment.

\textbf{Topology.} In Mininet, we will construct a network topology that includes a router, multiple switches, and host computers. The router is connected to all the switches and is responsible for forwarding traffic between different subnets. The switches, in turn, are connected to the hosts, enabling them to communicate within their respective subnets and interact with other devices on the network. To create various network topologies, we use two variables, \texttt{num\_switches} and \texttt{num\_hosts\_per\_subnet}, to control the setup. \texttt{num\_switches} defines how many switches are present in the network, while \texttt{num\_hosts\_per\_subnet} specifies how many hosts are connected to each switch. In our environment setup, the number of subnets and the number of hosts per subnet range from 2 to 4, making the topology dynamic and complex.

\begin{figure}[htb]
    \centering  
    \includegraphics[width=0.5\textwidth]{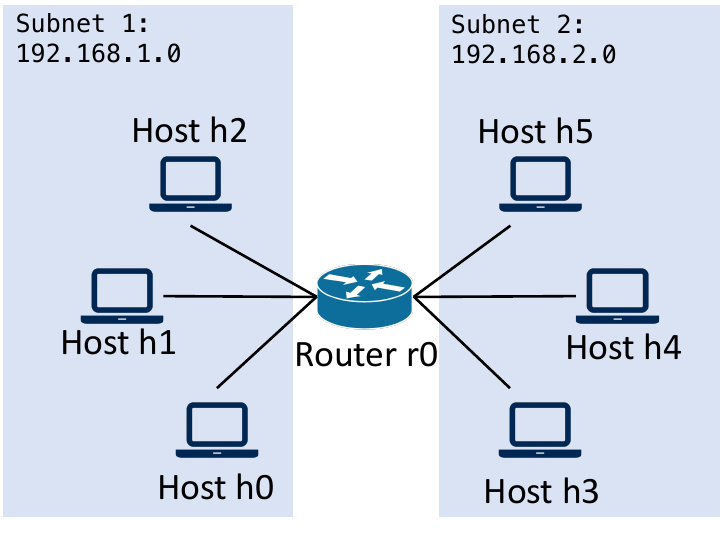} 
    % \vspace{-0.1cm}
    \caption{A simple version of routing network topology.}
    \label{fig:appendix:route}
\end{figure}

\subsection{Dynamic benchmark creation}

\textbf{Query Generation}
In the Topology section, we designed a network where all nodes are initially able to connect with one another. In our benchmark, each query represents a network state with an error, causing some nodes to be unable to communicate with certain others. We aim to have the LLM handle these error states and attempt to resolve the issues, ensuring that the hosts in the network can reconnect and communicate with each other.

To generate a problematic network, we start with a valid network configuration and apply a series of erroneous configuration commands to introduce faults. These faulty commands are selected from one of five basic categories of network errors, each with specific details that define the nature of the error. Importantly, within each error category, the details of the error can vary, leading to different manifestations of the same type of fault. This variability within error categories is what allows our benchmark to generate a diverse range of dynamic, distinct network errors, ensuring that the LLM is tested under various realistic fault conditions.

The five basic categories of network errors are as follows. We will describe each error type in detail, along with how the details within each error category can vary, ensuring a diverse set of dynamic network faults.

\textbf{Error Type 1: Disable Routing}

The error type \texttt{disable\_routing} is designed to simulate a failure in IP forwarding, which is crucial for routing packets between different subnets. In this error scenario, we disable IP forwarding using one of several methods. These methods are varied to generate diverse network configurations, making the benchmark dynamic and comprehensive. 

% \begin{tcolorbox}[colframe=gray!50!black, colback=yellow!10!white, coltitle=black, sharp corners=south, enhanced, breakable, width=\textwidth]
% \begin{lstlisting}[language=Python, 
%                    basicstyle=\ttfamily\small, % reduce font size
%                    keywordstyle=\bfseries\color{blue}, 
%                    commentstyle=\itshape\color{gray}, 
%                    stringstyle=\color{red!70!black}, 
%                    numbers=left, 
%                    numberstyle=\tiny\color{gray}, 
%                    frame=none, 
%                    showstringspaces=false, 
%                    breaklines=true, 
%                    breakatwhitespace=true]
% if errortype == "disable_routing":
%     method = errordetail.get("method", random.randint(1, 4))
%     print(f'*** Injecting error: Disabling IP forwarding using method {method}\n')
    
%     if method == 1:
%         print('*** Disabling IP forwarding using sysctl\n')
%         router.cmd('sysctl -w net.ipv4.ip_forward=0')
%     elif method == 2:
%         print('*** Disabling IP forwarding using iptables\n')
%         router.cmd('iptables -P FORWARD DROP')
%     elif method == 3:
%         print('*** Disabling IP forwarding using ip rule\n')
%         router.cmd('ip rule add prohibit pref 0')
%     elif method == 4:
%         print('*** Disabling IP forwarding for a specific subnet\n')
%         subnet_to_drop = errordetail.get("subnet", random.choice(subnets)[2])
%         router.cmd(f'iptables -A FORWARD -s {subnet_to_drop} -j DROP')
%     return
% \end{lstlisting}
% \end{tcolorbox}

There are four methods to disable IP forwarding. Method 1 uses \texttt{sysctl} to globally disable packet forwarding across all routes, while Method 2 utilizes \texttt{iptables} to drop all forwarded packets, affecting communication between subnets. Method 3 applies \texttt{ip rule} to prohibit forwarding based on specific rules, offering finer control. Method 4, also using \texttt{iptables}, drops traffic from a randomly selected subnet, creating a localized failure. These methods contribute to error diversity and allow the benchmark to simulate a wide range of network issues.

\textbf{Error Type 2: Disable Interface}

The error type \texttt{disable\_interface} simulates a failure by interfering with a specific network interface, which is crucial for communication between devices. In this error scenario, we have three methods to destroy a healthy network by disabling some interfaces.

% \begin{tcolorbox}[colframe=gray!50!black, colback=yellow!10!white, coltitle=black, sharp corners=south, enhanced, breakable, width=\textwidth]
% \begin{lstlisting}[language=Python, 
%                    basicstyle=\ttfamily\small, % reduce font size
%                    keywordstyle=\bfseries\color{blue}, 
%                    commentstyle=\itshape\color{gray}, 
%                    stringstyle=\color{red!70!black}, 
%                    numbers=left, 
%                    numberstyle=\tiny\color{gray}, 
%                    frame=none, 
%                    showstringspaces=false, 
%                    breaklines=true, 
%                    breakatwhitespace=true]
% if errortype == "disable_interface":
%     method = errordetail.get("method", random.randint(1, 3))
%     info(f'*** Injecting error: Interfering with interface {interface} using method {method}\n')
%     interface = errordetail["interface"]
%     if method == 1:
%         router.cmd(f'ifconfig {interface} down')
%     elif method == 2:
%         router.cmd(f'ip link set {interface} down')
%     elif method == 3:
%         router.cmd(f'ip link set {interface} mtu 68')
%     return
% \end{lstlisting}
% \end{tcolorbox}

Method 1 uses \texttt{ifconfig} to bring the interface down, resulting in a complete loss of connectivity through that interface. Method 2 utilizes \texttt{ip link} to achieve the same outcome of disabling the interface, but with a different network management tool. Method 3 changes the Maximum Transmission Unit (MTU) of the interface, which can cause packet fragmentation or loss if the MTU is set too low, leading to communication problems. These methods contribute to error diversity by offering various levels of disruption to simulate a wide range of interface failures.

\textbf{Error Type 3: Remove IP}

The error type \texttt{remove\_ip} is designed to simulate a failure by modifying or removing the IP address of a network interface. This can cause disruption in the network's ability to route packets, particularly if the interface's IP is essential for communication. The error is injected by using one of four methods to modify the IP address of the specified interface.

Method 1 flushes the IP using \texttt{ip addr flush}, making the interface unreachable. Method 2 assigns a random IP within the \texttt{10.0.0.0/24} subnet, which can cause conflicts or incorrect routing if the IP is already in use. Method 3 assigns a wrong subnet mask (e.g., \texttt{8}, \texttt{16}, \texttt{30}, \texttt{31}, or \texttt{32}), preventing communication due to misconfigured subnetting. Method 4 assigns a duplicate IP from another subnet, causing conflicts or routing issues, with a fallback IP used if no other subnets are available.

\textbf{Error Type 4: Drop Traffic to/From Subnet}

The error type \texttt{drop\_traffic\_to\_from\_subnet} simulates a failure by manipulating the traffic to or from a specific subnet. This can disrupt communication between the subnet and other network components. The error is injected by using one of four methods to modify the flow of traffic to/from the subnet.

Method 1 drops all incoming and outgoing traffic to/from the subnet using \texttt{iptables}. Method 2 actively rejects traffic to/from the subnet using \texttt{iptables} rules. Method 3 blocks ICMP traffic, preventing ping communication with the subnet. Method 4 introduces network delay using \texttt{tc netem}, adding latency to the traffic. These methods provide diverse ways to manipulate network traffic, enabling the benchmark to simulate various types of network disruptions.

\textbf{Error Type 5: Wrong Routing Table}

The error type \texttt{wrong\_routing\_table} simulates a failure by misconfiguring the routing table, which can disrupt the network's ability to properly forward traffic between subnets. The error is injected by modifying the routing table using one of four methods. Each method changes the routing behavior differently, enabling the benchmark to simulate various types of routing issues. 

Method 1 removes an existing route and adds a new route using a different interface. Method 2 adds a route with an incorrect gateway, potentially causing traffic to be misdirected. Method 3 adds a route with a very high metric, which makes it less preferred compared to other routes, possibly causing traffic delays or incorrect routing. Method 4 creates a routing loop by adding a route that forwards traffic to another subnet, which could lead to a loop of network traffic.

\textbf{Increasing Error Complexity: Combining Basic Errors}

The code above demonstrates the injection of basic network errors, but to generate more complex network states, we combine multiple basic errors. Specifically, we pair two different categories of errors and inject them sequentially into the network. This approach creates more intricate and realistic network failures, providing a better test for the LLM’s diagnostic capabilities. By combining multiple errors, the network's behavior becomes more complex, leading to a more challenging scenario for the LLM. The result of running the \texttt{pingall} command becomes more complicated, with more nodes failing to communicate. The LLM is faced with greater challenges, as it must analyze multiple potential causes, troubleshoot through various methods, and draw on a wider set of diagnostic skills. This makes the task significantly more difficult than diagnosing single errors in isolation.

The process of injecting two errors works as follows:

1. Single Error Injection: If only one error is to be injected (i.e., \texttt{errornumber} equals 1), the code only injects one single basic error. We use a function called \texttt{process\_single\_error} to handle this error type and inject it into the network.

% \begin{tcolorbox}[colframe=gray!50!black, colback=yellow!10!white, coltitle=black, sharp corners=south, enhanced, breakable, width=\textwidth]
% \begin{lstlisting}[language=Python, 
%                    basicstyle=\ttfamily\small, % reduce font size
%                    keywordstyle=\bfseries\color{blue}, 
%                    commentstyle=\itshape\color{gray}, 
%                    stringstyle=\color{red!70!black}, 
%                    numbers=left, 
%                    numberstyle=\tiny\color{gray}, 
%                    frame=none, 
%                    showstringspaces=false, 
%                    breaklines=true, 
%                    breakatwhitespace=true]
% if errornumber == 1:
%     print(f"Process {unique_id}: Injecting single error")
%     process_single_error(router, subnets, errortype, errordetail, unique_id)
% \end{lstlisting}
% \end{tcolorbox}
2. Multiple Error Injection: When injecting more than one error, \texttt{errortype} and \texttt{errordetail} are lists containing the error types and their corresponding details. Our benchmark uses a loop to pair elements from these lists and sequentially inject each error by calling \texttt{process\_single\_error} function for each pair.

% \begin{tcolorbox}[colframe=gray!50!black, colback=yellow!10!white, coltitle=black, sharp corners=south, enhanced, breakable, width=\textwidth]
% \begin{lstlisting}[language=Python, 
%                    basicstyle=\ttfamily\small, % reduce font size
%                    keywordstyle=\bfseries\color{blue}, 
%                    commentstyle=\itshape\color{gray}, 
%                    stringstyle=\color{red!70!black}, 
%                    numbers=left, 
%                    numberstyle=\tiny\color{gray}, 
%                    frame=none, 
%                    showstringspaces=false, 
%                    breaklines=true, 
%                    breakatwhitespace=true]
% else:
%     if isinstance(errortype, list) and isinstance(errordetail, list) and len(errortype) == errornumber and len(errordetail) == errornumber:
%         for et, ed in zip(errortype, errordetail):
%             process_single_error(router, subnets, et, ed, unique_id)
%     else:
%         print(f"Process {unique_id}: Error: For multiple error injection, errortype and errordetail must be lists of length equal to errornumber")
%         continue
% \end{lstlisting}
% \end{tcolorbox}

By combining different types of basic errors, the benchmark creates more complex network states that better simulate real-world network failures, providing the LLM with a more challenging test environment.

\subsection{LLM agents usage}
\textbf{Emulator.} In the Mininet simulator, an LLM is allowed to execute commands to retrieve information about network states and analyze the results of the \texttt{pingall} command. The \texttt{pingall} command sends ICMP echo requests (ping) from every host to each other in the network and reports the results. This is typically used to verify the connectivity between all nodes in a network, ensuring that the network is running as expected. A successful \texttt{pingall} output indicates that all hosts can communicate with each other, while failure may suggest network issues such as misconfigurations or connectivity problems.

The LLM can also utilize several types of commands to gather information about the network, diagnose issues, and propose solutions. 
To gather network information and diagnose issues, the LLM can use various commands such as \texttt{ifconfig}, \texttt{ip addr}, \texttt{ip link}, \texttt{ip route}. These commands allow the LLM to examine network configurations like IP address, network interfaces, routing tables. By analyzing the output from these commands, the LLM can identify problems such as incorrect configurations and suggest corrective actions to resolve the network issues. 
% \newpage
% \begin{tcolorbox}[colframe=gray!50!black, colback=yellow!10!white, sharp corners=south, enhanced, breakable]
% \textbf{Example Usage}
% \textbf{PingAll Results:}  \\
% Command: \texttt{PingAll} \\
% *** Fast Ping: testing ping reachability  

% h1 \texttt{->} X X X X X r0 \\
% h2 \texttt{->} X X X X X r0 \\
% h3 \texttt{->} X X X X X X \\
% h4 \texttt{->} X X X X X r0 \\
% h5 \texttt{->} X X X X X r0 \\
% h6 \texttt{->} X X X X X r0 \\
% r0 \texttt{->} h1 h2 X h4 h5 h6  

% *** Results: 76\% dropped (10/42 received)

% \textbf{Example feedback from Mininet:}  \\
% Command: \texttt{ip route} \\
% 192.168.1.0/24 dev r0-eth1 proto kernel scope link src 192.168.1.1 \\
% 192.168.2.0/24 dev r0-eth2 proto kernel scope link src 192.168.2.1 \\
% 192.168.3.0/24 dev r0-eth3 proto kernel scope link src 192.168.3.1 \\
% 192.168.4.0/24 dev r0-eth4 proto kernel scope link src 192.168.4.1 \\
% 192.168.5.0/24 dev r0-eth5 proto kernel scope link src 192.168.5.1 \\
% 192.168.6.0/24 dev r0-eth6 proto kernel scope link src 192.168.6.1  
% \end{tcolorbox}

\begin{PromptBlock}{Example Usage of Mininet Emulator}

\plabel{PingAll Results:}

\begin{lstlisting}[style=outputstyle]
Command: PingAll
*** Fast Ping: testing ping reachability

h1 -> X  X  X  X  X  r0
h2 -> X  X  X  X  X  r0
h3 -> X  X  X  X  X  X
h4 -> X  X  X  X  X  r0
h5 -> X  X  X  X  X  r0
h6 -> X  X  X  X  X  r0
r0 -> h1 h2 X  h4 h5 h6

*** Results: 76% dropped (10/42 received)
\end{lstlisting}

\plabel{Feedback from Mininet:}

\begin{lstlisting}[style=outputstyle]
Command: ip route
192.168.1.0/24 dev r0-eth1 proto kernel scope link src 192.168.1.1
192.168.2.0/24 dev r0-eth2 proto kernel scope link src 192.168.2.1
192.168.3.0/24 dev r0-eth3 proto kernel scope link src 192.168.3.1
192.168.4.0/24 dev r0-eth4 proto kernel scope link src 192.168.4.1
192.168.5.0/24 dev r0-eth5 proto kernel scope link src 192.168.5.1
192.168.6.0/24 dev r0-eth6 proto kernel scope link src 192.168.6.1
\end{lstlisting}

\end{PromptBlock}

% \textbf{Prompt Agent Creation}

% In this section, we use a base prompt that provides the fundamental information and instructions to the LLM. This prompt sets the task and guides the model in troubleshooting and diagnosing network issues in a Mininet environment. Below is the base prompt:

\textbf{Prompts.} We provide the initial prompt as below.

\begin{PromptBlock}{Prompt 2: Troubleshoot Routing Configuration}
\label{prompt:route}

You need to behave like a network engineer who finds the root cause of network issues and fixes them in a routing application.

There is a Mininet network with problems in the router \pcode{r0}, causing the network to be partially disconnected. Some nodes cannot successfully ping other nodes. Your task is to fix these issues so that the \pcode{pingall} result shows all connections are successful.

I recommend using diagnostic commands to gather information about the router and network to identify the cause of the problem. Once you have sufficient information and understand the root cause, provide commands to fix the issue.

\pimp{Important:} When implementing your solution, be careful not to disrupt existing connected edges --- your commands should \pkey{not} cause previously working connections to break.

Please provide your output in JSON format with the keys \pcode{machine} and \pcode{command}. You can only issue one command at a time as I can only execute commands sequentially.

\plabel{Constraints:}
\begin{itemize}[leftmargin=*, itemsep=2pt]
    \item The router's name may not be exactly \pcode{r0}. It may have a prefix (e.g., \pcode{p29\_r0}).
    \item The same applies to host names and interface names (e.g., \pcode{p29\_h1}, \pcode{p29\_r0-eth1}).
    \item The prefix could be anything (\pcode{p29}, \pcode{p30}, \pcode{p31}, etc.).
    \item Do \pkey{not} include \pcode{sudo} in your commands.
    \item You are \pkey{not} permitted to use the \pcode{vtysh} command.
    \item Do not use ping commands as the ping results are already provided to you.
\end{itemize}

I will provide you with the latest \pcode{PingAll()} feedback from the network along with your previous actions and their results to help you diagnose the problem.

\end{PromptBlock}

\subsection{Evaluation metric}
The performance evaluation of LLMs requires a comprehensive and diversified approach. A single metric only reflects the final success or failure of a command, but overlooks the intermediate steps in tasks such as network fault diagnosis, which often involve multiple iterations. For such tasks, it is essential to consider not only the end result, but also the necessity and efficiency of each intermediate step, which can be assessed through the number of iterations. Furthermore, evaluating whether these intermediate actions affect overall performance or even cause potential damage to the network is crucial as this represents a safety issue. By considering multiple dimensions of evaluation, we can gain a more holistic understanding of LLMs' capabilities, address their limitations, and ensure that they are safe, reliable, and effective for real-world deployment. Here are the evaluation metrics we considered.

\begin{itemize}
[topsep=1pt,itemsep=0ex,parsep=1ex, leftmargin=*]
    \item \textbf{Correctness:} This metric focuses on whether the final command execution results in a successful outcome, such as whether a \texttt{pingall} command succeeds. It reflects the accuracy of the LLM's final action in diagnosing and resolving faults. A high correctness rate indicates that the LLM reliably produces solutions that lead to successful outcomes without errors. 
    
    \item \textbf{Safety:} Safety is assessed by examining the LLM’s ability to preserve network stability during diagnosis and resolution. We expect the LLM to avoid issuing commands randomly, as arbitrary configuration changes could disrupt the network. In a real-world scenario, such actions could result in severe consequences, such as network downtime. Therefore, we aim to evaluate whether the LLM issues commands responsibly, ensuring it gathers sufficient information before taking action. If a command is issued to configure the network but fails to resolve the issue, the LLM will be penalized with a lower score, reflecting the unsafe nature of its response.

    \item \textbf{Latency:} This metric measures the number of iterations the LLM requires to reach a successful resolution. Fewer iterations to resolve the issue indicate that the LLM is more efficient in troubleshooting, as it can pinpoint the root cause of the problem accurately and provide the appropriate commands. An LLM with fewer iterations demonstrates higher efficiency in resolving network issues, leading to quicker resolutions and reduced network downtime. This improved efficiency is crucial for minimizing disruption, as it shows that the LLM is able to address network problems with precision and speed, ensuring better overall performance and reliability.
    
\end{itemize}

\section{Details for Microservice Policy Deployment Troubleshooting}
\label{appendix:k8s}
The third application focuses on troubleshooting tasks in Kubernetes and microservice policy configurations. Kubernetes (K8s), as the dominant orchestration platform, efficiently manages microservices by automating deployment, scaling, and orchestration, which significantly improves service scalability, resilience, and agility. However, misconfigured policies or faulty service deployments in Kubernetes clusters can lead to significant operational issues, including service downtime, performance bottlenecks, or critical security vulnerabilities. 

To evaluate whether LLM agents can autonomously identify and resolve Kubernetes network policy misconfigurations, we use Google's publicly available microservice benchmark as an emulator~\cite{google-k8s-demo}. Specifically, we leverage the Kubernetes-based microservice application composed of 11 services (e.g., frontend, checkout, payment) communicating via gRPC, secured by 13 distinct network policies that define permissible interactions among nodes and their ports. We dynamically generate realistic troubleshooting scenarios by injecting various types of network-policy misconfigurations. For each scenario, the LLM agent autonomously investigates the system, diagnoses the root cause, and attempts corrections to restore intended network connectivity. The LLM agent is deemed to have solved the query if it restores node communication, verified through connectivity tests, demonstrating its ability to automate Kubernetes configuration troubleshooting.

\subsection{Application environment}
This benchmark is based on Online Boutique, a cloud-first microservices demo application. Online Boutique is a web-based e-commerce platform where users can browse products, add them to their cart, and complete purchases. For our simulation, we use a local Kubernetes cluster to emulate the Online Boutique environment, providing an ideal setup for testing Kubernetes network policies in a real-world, cloud-native scenario.
\begin{figure}[htb]
    \centering  
    \includegraphics[width=0.65\textwidth]{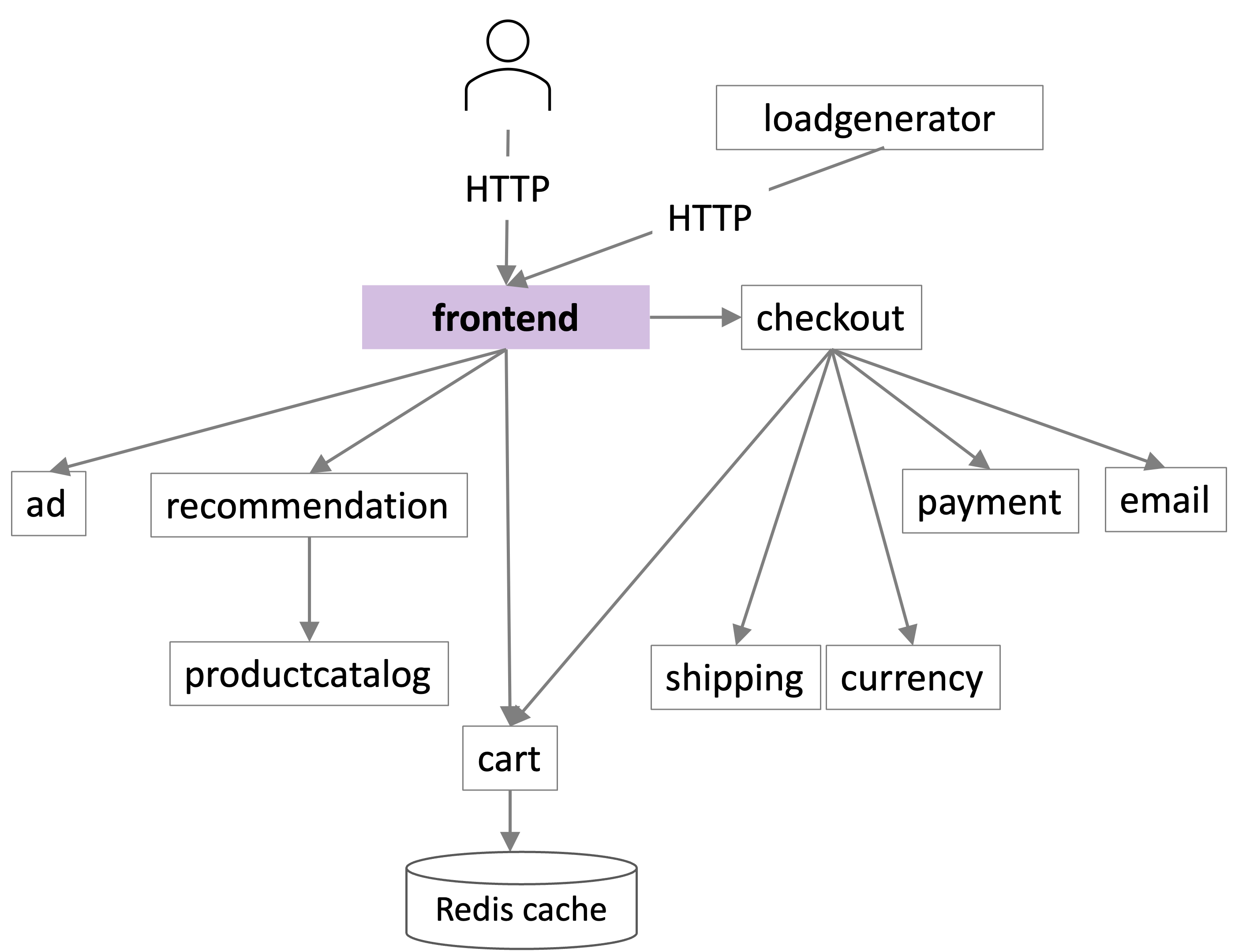} 
    % \vspace{-0.1cm}
    \caption{Structure of Google's microservice benchmark.}
    \label{fig:appendix:malt}
\end{figure}

The diagram above illustrates the architecture of the Online Boutique application, which consists of several microservices. In the diagram, each service is represented by a node, and the arrows indicate one-way communication between them. The one-way access is crucial for ensuring security and maintaining the integrity of the system. By restricting communication to only one direction, we can better control the flow of data and prevent unauthorized access or interactions between microservices, which is important for preserving the confidentiality and reliability of each service.

Our benchmark is designed to intentionally disrupt this structure by causing failures in the communication between the microservices. We will inject network misconfigurations into the cluster to simulate faults, which could lead to unauthorized access or prevent legitimate nodes from accessing the required services. The faulty cluster will then be provided to an LLM, allowing it to autonomously identify the root cause of the issues and take corrective actions, such as restoring proper access controls or resolving connectivity problems. This process will test the LLM’s ability to handle complex network policies and restore the system to its desired state.

\subsection{Dynamic benchmark creation}
In our benchmark design, we focus on error injection within a K8s cluster, specifically targeting the network configuration layer. To emulate real-world misconfigurations and generate diagnostic queries, we systematically inject faults by automatically modifying the network configuration YAML files. Our methodology is grounded on five fundamental types of basic network errors, which can also be added sequentially to increase complexity. These basic errors involve modifications to the ingress and egress rules, protocols, and ports within the Kubernetes network policies. By introducing such errors, we can observe how network disruptions manifest and how the LLM reacts to misconfigurations, which are critical for evaluating LLM's fault detection and correction abilities.

\textbf{Basic Network Policy Errors.} We generate examples of wrong policies deployed in the network, as follows.

\begin{itemize}[leftmargin=*, topsep=2pt, itemsep=1pt, parsep=0pt, partopsep=0pt]
    \item \textbf{Add Ingress Rule:} This error type involves adding new ingress rules to the network policy, which allows inbound traffic from external or unauthorized sources. In the YAML configuration, this would involve adding a new \texttt{from} rule under the \texttt{ingress} section.
    \begin{verbatim}
    ingress:
      - from:
        - podSelector:
            matchLabels:
              app: frontend
    \end{verbatim}
    \textbf{Potential Impact:} Adding an ingress rule with an external source could compromise the security boundaries of the system, allowing unauthorized access to internal services, which might lead to security breaches.

    \item \textbf{Add Egress Rule:} This error type adds a new egress rule to allow internal services to communicate with external services or networks, potentially violating security protocols. The YAML change typically involves adding a new \texttt{to} rule under the \texttt{egress} section.
    \begin{verbatim}
    egress:
      - to:
        - podSelector:
            matchLabels:
              app: frontend
    \end{verbatim}
    \textbf{Potential Impact:} Adding an egress rule might cause unauthorized data leaks or expose the system to connections with external entities. It can allow sensitive data to flow out of the network, risking privacy or system integrity.

    \item \textbf{Remove Ingress Rule:} Deleting an ingress rule can block necessary traffic from reaching internal services. In the YAML configuration, this would involve removing an existing ingress rule under the \texttt{ingress} section.
    \begin{verbatim}
    ingress:
      # Removed rule here
      - from:
        - podSelector:
            matchLabels:
              app: frontend
    \end{verbatim}
    \textbf{Potential Impact:} By removing an ingress rule, valid traffic from authorized services may be blocked, leading to service downtime or failure to respond to incoming requests. This can reduce system availability and affect service reliability.

    \item \textbf{Change Protocol:} Changing the communication protocol (e.g., from TCP to UDP or vice versa) can disrupt inter-service communication. The YAML configuration might be modified by altering the \texttt{protocol} field under the \texttt{ports} section.
    \begin{verbatim}
    ports:
      - port: 9555
        protocol: UDP  % Changed from TCP to UDP
    \end{verbatim}
    \textbf{Potential Impact:} If services depend on a specific protocol for communication, changing it can result in connectivity issues. Services may fail to establish connections, causing service interruptions and degraded performance.

    \item \textbf{Change Port:} Modifying the port number used for communication between services can lead to issues such as services becoming unreachable or port conflicts. This would be represented by modifying the \texttt{port} value under the \texttt{ports} section of the YAML configuration.
    \begin{verbatim}
    ports:
      - port: 8080  % Changed from 9555 to 8080
        protocol: TCP
    \end{verbatim}
    \textbf{Potential Impact:} Changing the port configuration can make services unreachable if other services still expect the old port. Port conflicts can arise if another service is already using the new port, resulting in failed connections and network instability.
\end{itemize}

\textbf{Increasing Complexity through Sequential Error Injection.}  
To increase the difficulty and complexity of network misconfigurations, we adopt a sequential error injection strategy. This approach involves identifying and modifying two separate YAML files, each representing different aspects of the network configuration. By injecting different types of errors into each file, we create a more intricate and challenging scenario for the Kubernetes cluster. Sequential injection of multiple error types forces LLMs to handle progressively more complex network misconfigurations, which can test the LLM’s ability to diagnose and resolve issues more effectively.

\subsection{LLM agents usage}  

\textbf{Emulator.}
In this section, we describe the setup of an emulation platform designed to facilitate interactions between LLMs and a K8s cluster. The goal of this environment is to create a testing ground where LLMs can be used to diagnose and resolve network misconfigurations within the K8s cluster. The environment allows us to introduce various network issues, test LLM-based troubleshooting techniques, and evaluate the efficiency and accuracy of LLM interventions.
 
For our simulation, we utilized the kind simulator, a tool designed for running local Kubernetes clusters using Docker container "nodes". Using kind, we can simulate an entire Kubernetes cluster on a single virtual machine (VM), which provides a lightweight and easy-to-manage environment for our tests. In addition, we deploy Google's microservices demo locally, benefiting from the compact nature of the setup and the simplicity of management. Besides, the Kind simulator fully replicates the behavior of a real Kubernetes cluster, ensuring that the environment that we provide to the LLM is consistent with production-like conditions. This allows us to present the LLM with an environment that mimics real-world Kubernetes clusters, enabling accurate and reliable testing of its fault detection and resolution capabilities.

Once the local K8s cluster was successfully set up, we can inject errors to simulate network misconfigurations and connectivity issues. After the error injection, we have a function to collect connectivity failures due to configuration errors between nodes. This information is then provided to the LLM, which will help it analyze the current network state and generate a Kubernetes command to resolve the identified issue. The command is executed within the environment, and the updated connectivity status is subsequently fed back to the LLM. Equipped with this new information, the LLM interacts with the system, refining its diagnosis, and issuing additional commands if necessary. This iterative process continues until the network issues are resolved, enabling the LLM to effectively troubleshoot and fix network problems. Through this approach, the benchmark assesses the diagnostic accuracy, problem-solving efficiency, and adaptability of the LLM in dynamic network environments, providing a comprehensive assessment of its performance in real-world scenarios. The following provides the implementation details for the network status check and the interaction with LLMs.

\textbf{Network Status Check.}  
We have implemented a function that automatically checks the network connectivity between nodes in the K8s cluster to identify any discrepancies between the actual communication and the expected connectivity. 

At the beginning of the network status check, a debug container (which will be reused for the entire benchmarking process) is created for each pod, containing basic network tools necessary for connectivity testing. During the network status testing process, we access the debug container and use the `nc` (Netcat) command to test whether a pod can communicate with other pods within the cluster. The results of these tests are then provided to the LLM in the form of a mismatch report, detailing which nodes have communication issues that differ from the expected connectivity.

% \begin{tcolorbox}[colframe=gray!50!black, colback=yellow!10!white, sharp corners=south, enhanced, breakable]
% \textbf{Example Mismatch}

% Mismatch Summary:

% frontend → adservice:9555 (Expected: True, Actual: False)

% frontend → cartservice:7070 (Expected: False, Actual: True)

% \textbf{Explanation:}

% In the above mismatch results, the connectivity between pods is compared against the expected behavior. The first mismatch indicates that the `frontend` pod was expected to communicate with the `adservice` pod on port 9555, but the actual connectivity failed (Expected: True, Actual: False). In contrast, the second mismatch shows that the `frontend` pod was expected not to communicate with the `cartservice` pod on port 7070, but the actual connection was established (Expected: False, Actual: True). 
% \end{tcolorbox}
\begin{PromptBlock}{Example Mismatch Output from K8s Emulator}

\plabel{Mismatch Summary:}

\begin{lstlisting}[style=outputstyle]
frontend -> adservice:9555   (Expected: True,  Actual: False)
frontend -> cartservice:7070 (Expected: False, Actual: True)
\end{lstlisting}

\plabel{Explanation:}

In the above mismatch results, the connectivity between pods is compared against the expected behavior.
\begin{itemize}[leftmargin=*, itemsep=2pt]
    \item The first mismatch indicates that the \pcode{frontend} pod was expected to communicate with \pcode{adservice} on port \pcode{9555}, but the actual connectivity failed (\pcode{Expected: True, Actual: False}).
    \item The second mismatch shows that \pcode{frontend} was expected \pkey{not} to communicate with \pcode{cartservice} on port \pcode{7070}, but the connection was established (\pcode{Expected: False, Actual: True}).
\end{itemize}

\end{PromptBlock}

\textbf{LLM Interaction.}  
To enable effective interaction between the LLM and the K8s cluster, we allow the LLM to use Kubernetes commands to troubleshoot and resolve network issues. When the LLM analyzes the network status and identifies potential problems, it generates Kubernetes commands as output. These commands are then extracted from the LLM’s response and executed on the K8s cluster.

To run the generated Kubernetes commands, we use Python’s `subprocess` module, which allows us to programmatically execute shell commands. The command’s output, including any errors or status messages, is captured and stored for further analysis. This enables us to monitor the LLM’s actions and track the results of the commands it issues.

By interacting with the K8s cluster in this way, the LLM is able to perform network troubleshooting tasks, such as diagnosing misconfigurations and resolving connectivity issues. The model uses the output from each command to refine its diagnosis, iterating on its approach if necessary. Through this process, the LLM can progressively identify the root causes of network problems and issue further corrective commands until the issues are resolved, completing the troubleshooting process.

\textbf{Prompts.} We provide the initial prompt as below.
\begin{PromptBlock}{Prompt 3: Troubleshoot Microservice Policy }

You need to behave like a network engineer who can find the root cause of network policy deployment issues and fix them in a microservices architecture.

\plabel{Service Communication Topology:}

Our microservices architecture contains the following services and desired communication relationships:
\begin{itemize}[leftmargin=*, itemsep=2pt]
    \item \pcode{user} and \pcode{loadgenerator} can access the \pcode{frontend} service via HTTP.
    \item \pcode{frontend} communicates with: \pcode{checkout}, \pcode{ad}, \pcode{recommendation}, \pcode{productcatalog}, \pcode{cart}, \pcode{shipping}, \pcode{currency}, \pcode{payment}, and \pcode{email}.
    \item \pcode{checkout} further communicates with \pcode{payment}, \pcode{shipping}, \pcode{email}, and \pcode{currency}.
    \item \pcode{recommendation} communicates with \pcode{productcatalog}.
    \item \pcode{cart} communicates with the \pcode{Redis} cache for storing cart data.
\end{itemize}

Your task is to inspect the current network policies and verify if they meet the described communication patterns. If there are any \pkey{mismatches}, you should fix them.

\plabel{Interaction Protocol:}
\begin{itemize}[leftmargin=*, itemsep=2pt]
    \item Provide \pkey{one command at a time} to check connectivity or node accessibility.
    \item Each turn, I will provide your previous commands and their corresponding outputs.
    \item I will also provide the current connectivity status, including any mismatches between the expected and actual connectivity.
    \item Use this information to identify and fix misconfigurations step-by-step.
\end{itemize}

\end{PromptBlock}

\subsection{Evaluation Metric.}

The performance evaluation of LLMs for K8s cluster troubleshooting requires a comprehensive and multidimensional approach. A single metric only reflects the final success or failure of a command but overlooks the intermediate steps in tasks like network fault diagnosis, which often involve multiple iterations. For tasks in a K8s cluster, it's important to evaluate not only the final result but also the necessity and efficiency of each intermediate action, which can be assessed by the number of iterations. Additionally, it’s crucial to evaluate whether these intermediate actions impact overall performance or cause potential damage to the cluster, as this poses a safety risk. By considering these different dimensions, we can gain a more thorough understanding of the LLM's capabilities, identify its limitations, and ensure that it is safe, reliable, and effective in real-world K8s deployments. The evaluation metrics we considered are as follows:

\begin{itemize}[topsep=1pt,itemsep=0ex,parsep=1ex, leftmargin=*]
    \item \textbf{Correctness:} This metric focuses on whether the final command execution results in a successful outcome. We will use the network status check in the previous section to find if the K8s cluster gets to the expected state with the LLM's help. It reflects the accuracy of the LLM's final action in diagnosing and resolving network faults in a K8s environment. A high correctness rate indicates that the LLM reliably produces solutions that lead to successful outcomes without errors. 
    
    \item \textbf{Safety:} Safety is measured by evaluating the LLM’s ability to maintain the K8s cluster’s stability during the diagnosis and resolution process. This includes monitoring network status during the LLM troubleshooting process to see if the LLM's command will destroy original connectivity. For instance, an increase in pod failures or a loss of network connectivity could indicate that the LLM’s commands are destabilizing the existing system state, potentially leading to larger failures in production environments.

    \item \textbf{Latency:} This metric measures the number of iterations the LLM requires to reach a successful resolution of the issue within a K8s cluster. Fewer iterations to resolve the issue indicate that the LLM is more efficient in troubleshooting, as it can more accurately pinpoint the root cause of the problem and generate the appropriate Kubernetes commands. An LLM that reaches a solution in fewer iterations demonstrates higher efficiency in addressing K8s network issues, leading to quicker resolutions and reduced downtime. This efficiency is essential in a production environment, where minimizing network disruption is critical to maintaining service availability and system reliability.
\end{itemize}
\newpage
\begin{table}
\centering
\resizebox{1.0\columnwidth}{!}{
\begin{tblr}{
  column{2} = {c},
  column{3} = {c},
  cell{2}{1} = {r=4}{},
  cell{6}{1} = {r=3}{},
  cell{9}{1} = {r=3}{},
  hline{1-2,6,9,12} = {1-3}{},
}
                 & \textbf{Query Example} & \textbf{Action Label} &  &  &  &  \\
\textbf{Level 1} & Remove node \texttt{A}. List the direct child nodes of \texttt{A}'s parent in the updated graph.                & RM               &  &  &  &  \\
                 & Rank all child nodes of \texttt{CONTROL\_DOMAIN} under node \texttt{B} based on capacity.                & R                 &  &  &  &  \\
                 & List all children of \texttt{B}. Return all children names.                & LI                 &  &  &  &  \\
                 & Add a new node \texttt{P} with \texttt{PORT} type to node \texttt{C}.                & A                  &  &  &  &  \\
\textbf{Level 2} & Remove node \texttt{D}. Count nodes with \texttt{PORT} under \texttt{D}'s parent in the updated graph.
                & RM-C         &  &  &  &  \\
                 & Remove node \texttt{A} from the graph. List the direct child nodes of \texttt{A}'s parent.
                & RM-LI          &  &  &  &  \\
                 & Remove node \texttt{A}. Rank the child nodes of \texttt{A}'s parent based on the total bandwidth.
                & RM-R          &  &  &  &  \\
\textbf{Level 3} & Add node \texttt{E} to node \texttt{F}. Count the number of \texttt{PACKET\_SWITCH} under \texttt{F}.
                & A-C            &  &  &  &  \\
                 & Add node \texttt{E} to node \texttt{G}. List the direct child nodes of \texttt{G} in the updated graph.
                & A-LI             &  &  &  &  \\
                 & Add node \texttt{E} to node \texttt{H}. Rank the child nodes of \texttt{H} based on the total bandwidth.
                & A-R             &  &  &  &  
\end{tblr}
}
\caption{Action and complexity level details for queries in CP.}
\label{tab:append:query_details:malt}
\end{table}
\begin{table}
\centering
\makebox[\textwidth][c]{%
\resizebox{0.9\columnwidth}{!}{
\begin{tblr}{
  column{2} = {c},
  column{3} = {c},
  cell{2}{1} = {r=5}{},
  cell{7}{1} = {r=7}{},
  cell{14}{1} = {r=3}{},
  hline{1-2,7,14,17} = {1-3}{},
}
                 & \textbf{Error Details}                                & \textbf{Error Label} &  &  &  &  \\
\textbf{Level 1} & disable\_routing                                      & DR                          &  &  &  &  \\
                 & disable\_interface                                    & DI                          &  &  &  &  \\
                 & remove\_ip                                            & RI                          &  &  &  &  \\
                 & drop\_traffic\_to\_from\_subnet                       & DT                          &  &  &  &  \\
                 & wrong\_routing\_table                                 & WR                          &  &  &  &  \\
\textbf{Level 2 } & disable\_routing + disable\_interface                   & DR+DI                       &  &  &  &  \\
                 & disable\_routing + remove\_ip                           & DR+RI                       &  &  &  &  \\
                 & disable\_routing + drop\_traffic\_to\_from\_subnet      & DR+DT                       &  &  &  &  \\
                 & disable\_routing + wrong\_routing\_table                & DR+WR                       &  &  &  &  \\
                 & remove\_ip + wrong\_routing\_table                      & RI+WR                       &  &  &  &  \\
                 & drop\_traffic\_to\_from\_subnet + wrong\_routing\_table & DT+WR                       &  &  &  &  \\
                 & disable\_interface + drop\_traffic\_to\_from\_subnet    & DI+DT                       &  &  &  &  \\
\textbf{Level 3 } & disable\_interface + wrong\_routing\_table              & DI+WR                       &  &  &  &  \\
                 & remove\_ip + drop\_traffic\_to\_from\_subnet            & RI+DT                       &  &  &  &  \\
                 & disable\_interface + remove\_ip                         & DI+RI                       &  &  &  &  
\end{tblr}
}}
\centering
\caption{Error and complexity level details for queries in Routing.}
\label{tab:append:query_details:k8s}
\end{table}
\begin{table}
\centering
\resizebox{0.75\textwidth}{!}{
\centering
\begin{tblr}{
  column{2} = {c},
  column{3} = {c},
  cell{2}{1} = {r=5}{},
  cell{7}{1} = {r=6}{},
  cell{13}{1} = {r=4}{},
  hline{1-2,7,13,17} = {1-3}{},
}
                 & \textbf{Error Details}           & \textbf{Error Label} &  &  &  &  \\
\textbf{Level 1} & remove\_ingress                  & RI                          &  &  &  &  \\
                 & add\_ingress                     & AI                          &  &  &  &  \\
                 & change\_port                     & CP                          &  &  &  &  \\
                 & change\_protocol                 & CPR                         &  &  &  &  \\
                 & add\_egress                      & AE                          &  &  &  &  \\
\textbf{Level 2 } & remove\_ingress + add\_ingress     & RI+AI                       &  &  &  &  \\
                 & remove\_ingress + change\_port     & RI+CP                       &  &  &  &  \\
                 & remove\_ingress + change\_protocol & RI+CPR                      &  &  &  &  \\
                 & add\_ingress + change\_port        & AI+CP                       &  &  &  &  \\
                 & add\_ingress + change\_protocol    & AI+CPR                      &  &  &  &  \\
                 & change\_port + change\_protocol    & CP+CPR                      &  &  &  &  \\
\textbf{Level 3 } & change\_port + add\_egress         & CP+AE                       &  &  &  &  \\
                 & change\_protocol + add\_egress     & CPR+AE                      &  &  &  &  \\
                 & remove\_ingress + add\_egress      & RI+AE                       &  &  &  &  \\
                 & add\_ingress + add\_egress         & AI+AE                       &  &  &  &  
\end{tblr}
}
\caption{Error and complexity level details for queries in K8s.}
\label{tab:append:query_details:route}
\end{table}

\newpage

\end{document}